\documentclass[fleqn,usenatbib]{mnras}

\usepackage{graphicx}      
\usepackage{float}
\usepackage{amsmath}
\usepackage{amssymb}
\usepackage{threeparttable}
\usepackage{xcolor}
\usepackage{orcidlink}
\usepackage{url}            

\DeclareGraphicsExtensions{.pdf,.png,.jpg,.jpeg}

\DeclareRobustCommand{\VAN}[3]{#2}
\let\VANthebibliography\thebibliography
\def\thebibliography{\DeclareRobustCommand{\VAN}[3]{##3}\VANthebibliography}

\graphicspath{{figures/}}

\usepackage{newtxtext,newtxmath}   
\usepackage[T1]{fontenc}           

\usepackage{natbib}

\title[Search for scattering substructure in AGNs]{Search for refractive substructure in active galactic nuclei using ground-based VLBI observations}

\author[Koryukova et al.]{
\parbox{\textwidth}{
T.~A. Koryukova,$^{1}$\thanks{E-mail: tatyana.koryukova@gmail.com}\orcidlink{0000-0001-8347-7880}
A.~V. Plavin,$^{2}\orcidlink{0000-0003-2914-8554}$
A.~B. Pushkarev,$^{3,1}$\orcidlink{0000-0002-9702-2307}
Y.~Y. Kovalev$^{4}$\orcidlink{0000-0001-9303-3263}
}
\vspace{0.4cm}\\
\parbox{\textwidth}{
$^1$Lebedev Physical Institute of the Russian Academy of Sciences, Leninsky prospekt 53, 119991, Moscow, Russia\\
$^2$Black Hole Initiative, Harvard University, 20 Garden St, Cambridge, MA 02138, USA\\
$^3$Crimean Astrophysical Observatory, Nauchny 298409, Crimea, Russia\\
$^4$Max-Planck-Institut f\"ur Radioastronomie, Auf dem H\"ugel 69, 53121 Bonn, Germany
}
}

\date{Accepted 2026 ZZZ. Received 2026 YYY ; in original form 2026 August 28}

\pubyear{2026}

\begin{document}
\label{firstpage}
\pagerange{\pageref{firstpage}--\pageref{lastpage}}
\maketitle

\begin{abstract}
We present the results of the first systematic search for refractive scattering substructure in active galactic nuclei (AGN) using ground-based very long baseline interferometry (VLBI) data. We selected 1074 AGNs showing a steep frequency dependence of their angular sizes and compared the observed long-baseline visibility amplitudes with the predictions of the refractive substructure model for scatter-broadened images. Nine sources showed consistency with the expected level of refractive noise, including the quasar TXS~2005$+$403 (J2007$+$4029), which confirms previously reported results for this source. Remarkably, the majority of these AGNs are located in the Cygnus region -- one of the key ``hotspots'' of strong Galactic scattering. For five selected AGNs, we identified at least two epochs with sufficiently high data quality showing long-baseline detections, suggesting the presence of scattering substructure. We showed that refractive scattering substructure in AGN is a rare effect, requiring high sensitivity, sufficient source brightness, and lines of sight with particularly strong turbulence in the interstellar medium. Nevertheless, AGNs can serve as powerful probes of the turbulent interstellar medium. We discuss strategies for improving the sensitivity of VLBI observations and identify the most promising sources for reliable detection and investigation of refractive substructure in future observations.
\end{abstract}



\section{Introduction}
\label{s:introduction}

The ionized interstellar medium (ISM) of the Galaxy contains electron density fluctuations that scatter radio emission from compact extragalactic sources such as active galactic nuclei \citep[e.g.,][]{Pushkarev15, Koryukova2022}. These fluctuations, described by the Kolmogorov spectrum of turbulence \citep{Kolmogorov41}, act as an inhomogeneous ``screen'' that distorts the source image. In very long baseline interferometry (VLBI), this manifests in two primary regimes. Diffractive scattering on small-scale inhomogeneities leads to angular broadening of the source size \citep{Duffett_Smith1976, Rickett1990}. On larger turbulence scales exceeding the size of the Fresnel zone, scattering produces fine-scale substructure within the broadened image \citep{Narayan1989, Goodman1989, Gwinn2014, JohnsonGwinn2015, Johnson2016, SgrA_Johnson18, Issaoun2019}. In VLBI observations, this substructure appears as a persistent non-zero signal on long baselines, where the correlated flux is expected to be completely suppressed according to the predictions of the diffractive scattering model. It is known that scattering substructure can mimic compact intrinsic source components and lead to systematic overestimates of brightness temperature, which is critical for the correct interpretation of physical parameters of relativistic AGN jets \citep{Johnson2016, Kovalev2016, Pilipenko2018}.

Direct detection of refractive substructure in AGN has long been difficult, mainly due to two reasons: limited sensitivity of the observations \citep[e.g.,][]{JohnsonGwinn2015}, and the degeneracy between scattering effects and the intrinsic compact structure of the source \citep[e.g.,][]{Pilipenko2018}. Nevertheless, observations with the ground-space interferometer RadioAstron have shown signs of such phenomena for the sources 3C\,273 \citep{Johnson2016} and B0529$+$483 \citep{Pilipenko2018}. A breakthrough in this direction was the work of \cite{Plavin2026}, which presented the first robust detection of refractive substructure in an AGN using ground-based VLBI observations. The object of study was the blazar TXS~2005$+$403 (J2007$+$4029), whose line of sight passes through the Cygnus region. The authors showed that the visibility amplitudes observed on long baselines in the frequency range 1--5~GHz cannot be explained by angular broadening due to diffractive scattering of the radio emission. It was shown that diffractive broadening dominates on short baselines, while refractive substructure dominates the signal on long baselines and is detected on timescales of at least a decade.

The aim of this work is a systematic search for new sources exhibiting signatures of scattering substructure, using data from AGNs observed in ground-based VLBI programs. We used all archival data available in the Astrogeo image collection  \citep{2025ApJS..276...38P}. Our approach is based on a direct comparison of measured visibility amplitudes on long baselines with the level of refractive noise predicted by the theoretical model of \cite{JohnsonGwinn2015} (hereafter JG15) for a Kolmogorov turbulence spectrum with $\alpha=5/3$ and an inner turbulence scale $r_{in}=1000$~km \citep{JohnsonGwinn2015, Plavin2026, Pushkarev2026}. Observations indicate that the inner scale is typically between 100 and 1000~km, although its exact value depends on the specific line of sight \citep[e.g.,][]{Spangler1990_1, Rickett2009, Johnson2021}. In contrast to the single-source study of \cite{Plavin2026}, our work is based on an automated search for long-baseline signals in all available sources and their observing epochs. This can potentially reveal new sources exhibiting similar effects, provide a statistical estimate of the prevalence of refractive scattering among AGNs, and establish a connection between this phenomenon and the parameters of the interstellar medium along the line of sight.

\section{Archival VLBI data}
\label{s:data}

Our analysis is based on VLBI observations of active galactic nuclei jets in the frequency range from 1.4 to 8.7~GHz, compiled in the Astrogeo database. The observational data include results from long-term survey programs and targeted observations carried out from 1994 to 2025. They are based on geodetic VLBI observations \citep{2009JGeod..83..859P,2012A&A...544A..34P,2012ApJ...758...84P}, VLBA (Very Long Baseline Array) calibrator surveys \citep{2002ApJS..141...13B,2003AJ....126.2562F,2005AJ....129.1163P,2006AJ....131.1872P,2007AJ....133.1236K,2008AJ....136..580P}, as well as observations from other networks including EVN (European VLBI Network), LBA (Long Baseline Array), and GMVA (Global millimeter VLBI array) \citep{2007ApJ...658..203H,2008AJ....136..159L,2011AJ....142...35P,2011MNRAS.414.2528P,2011AJ....142..105P,2012MNRAS.419.1097P,2013AJ....146....5P,2015ApJS..217....4S,2017ApJS..230...13S,2017ApJ...846...98J,2019MNRAS.485...88P,2019A&A...622A..92N,2021AJ....161...14P,2021AJ....161...88P}. In the selected frequency range, a total of 19\,691 sources were observed over the entire period. The total number of source-epochs is 95\,599. The dataset, which spans a wide frequency range and numerous lines of sight through the Galaxy, offers a unique opportunity for identifying strongly scattered sources.

We fit the observed brightness distribution of each source with a single Gaussian component. The choice of a single-component model is justified by the fact that at the frequencies used in this study ($\le8$~GHz), complex jet structure is typically completely suppressed by diffractive angular broadening and remains unresolved if scattering is indeed significant \citep{Koryukova2023, Plavin2026}. To estimate the angular sizes of the sources, we used directly measured interferometric visibilities rather than reconstructed images. This approach avoids the artifacts inevitably introduced during the deconvolution of VLBI maps and preserves the maximum accuracy when modeling the source structure. For the fitting procedure, we employed the Nested Sampling algorithm, which allows the process to be fully automated and independent of the initial guess. This approach ensures reproducibility of the results for thousands of sources in the sample. A detailed description and justification of the methodology for estimating the angular sizes of AGN jet components is presented in \cite{plavin2022}. We excluded from the analysis measurements for which the reconstructed angular size of the source does not exceed the effective resolution limit of the interferometer. As a result, approximately $23\%$ of the original data were discarded. The final sample comprises 13\,309 sources and 73\,410 individual observations.

\section{Methods}
\label{s:methods}

\subsection{Estimating the predicted level of refractive noise}
\label{ss:noise_level}

According to the JG15 model, even on baselines where the diffractive scattering profile is well below the thermal noise level, the visibility amplitude does not drop to zero completely. Instead, it fluctuates around a baseline-dependent mean level. This model defines the mean fluctuation level of the visibility amplitude $\sigma_\mathrm{ref}$, arising from refractive substructure in the scattered image, for given source and scattering screen parameters. Following the JG15 approach, the standard deviation of the normalized visibility amplitude at a projected baseline $\mathbf{b}$ can be estimated using the relation:

\begin{equation}
\begin{split}
\sigma_{\rm ref}(\mathbf{b}) = C \times \left( \frac{\lambda}{\lambda_0} \right) 
\left( \frac{|\mathbf{b}|}{10^5~{\rm km}} \right)^{-5/6} 
\left( \frac{\theta_{\rm scatt}}{\theta_{\rm scatt,0}} \right)^{5/6} \\
\times \left( \frac{\theta_{\rm img}}{\theta_{\rm img,0}} \right)^{-2} 
\left( \frac{D}{1~{\rm kpc}} \right)^{-1/6}
\end{split}
\label{eq:JG19}
\end{equation}

\noindent
where $\lambda$ is the observing wavelength; $\mathbf{b}$ is the projected baseline length in the source direction; $\theta_{\rm img}$ is the observed angular size of the source; and $\theta_{\rm scatt}$ is the angular broadening due to diffractive scattering. The normalization parameters $\lambda_0$, $\theta_{\rm scatt,0}$, $\theta_{\rm img,0}$ and the constant $C$ are defined separately for three wavelength ranges used in RadioAstron mission, as given in eq.~(19) in JG15 paper. We adopt the version of eq.~(19) with normalization parameters for the 6~cm wavelength range. The distance from the screen to the observer $D$ does not significantly affect the refractive noise level in this model, so we assume that the scattering material is located within the Milky Way at a typical distance of $D = 1$~kpc. Following JG15, we assume a Kolmogorov turbulence spectrum with $\alpha = 5/3$ and an inner scale $r_{\rm in} = 1000$~km. If the substructure contribution is significant, the detections are expected to fall within the predicted 95\% confidence interval of the Rayleigh distribution, defined as $[0.225, 2.72] \times \sigma_{\rm ref}(\mathbf{b})$. These lower and upper boundaries correspond to the 2.5\% and 97.5\% quantiles of the Rayleigh distribution, respectively. In this work, we do not attempt to constrain the turbulence screen properties $\alpha$ and $r_{\rm in}$, but rather perform a consistency check of the observed signal level with the predictions of the JG15 theory.

The key parameter in the model is the observed angular size of the source $\theta_{\rm img}$. To determine it, we use an approach based on modeling the interferometric visibilities with circular Gaussian component, as already mentioned in Section~\ref{s:data}. In this case, $\theta_{\rm img}$ is the full width at half maximum (FWHM) of the Gaussian distribution fitted to the observed visibility function. In cases where scattering dominates the observed structure, any intrinsic source morphology is effectively smoothed by the turbulent screen, and the visibility function can be well approximated by a single Gaussian component. Thus, the size of the resulting $\theta_{\rm img}$ contains information about the properties of the scattering screen along the line of sight.

\subsection{Validating long-baseline visibility measurements}
\label{ss:detections}

To determine which of the long-baseline visibilities in the database correspond to robust signal detections, we re-analyse the raw correlator output data from the VLBA archive\footnote{\url{https://data.nrao.edu}}, following the approach of \citet{Plavin2026}. The method is a barebones fringe search: for each scan, baseline, and frequency range, the visibilities are Fourier-transformed over time and frequency, and the resulting delay--fringe-rate spectrum is searched for a significant peak. The search covers the full available window: $1$--$8$~$\mu$s in delay (determined by the spectral channel spacing) and $0.1$--$5.0$~Hz in fringe rate (depending on the correlator integration time).
The difference from the standard VLBI data processing (e.g.  AIPS, \citealt{Greisen2003}) is that here detections are established in the most conservative way: each baseline is treated independently, the search runs on the raw correlator output before any processing steps. The probability that the delay--rate peak is spurious and noise-driven is estimated as:
\begin{equation}
P(\mathrm{peak} \mid \mathrm{noise}) \leq N e^{-\mathrm{SNR}^2 / 2},
\end{equation}
where $N$ is the number of search grid cells, SNR is signal-to-noise ratio. Oversampling introduces correlations between neighboring cells, so the actual number of independent values is smaller than $N$, making the above formula a conservative estimate of the false detection probability. We adopt a strict threshold of $P(\mathrm{peak} \mid \mathrm{noise}) < 5 \times 10^{-5}$, which corresponds to fewer than one expected spurious detection across all scans, baselines, and IFs (intermediate frequency) analysed in this work. Additional details of the method are discussed in \cite{Plavin2026}.

The primary way fringe search is performed is the \textit{combined} mode: adjacent IFs are combined into a single band before the search, providing an SNR gain proportional to $\sqrt{n}$, where $n$ is the number of IFs; this is the most sensitive mode, assuming that IFs are coherent in the telescope data. In the \textit{per-IF} mode, each IF is searched independently, enabling us to handle situations in cases where between-IF coherence is broken.

\section{Sample selection}
\label{s:sample}

\begin{figure}
    \centering
    \includegraphics[width=1\linewidth]{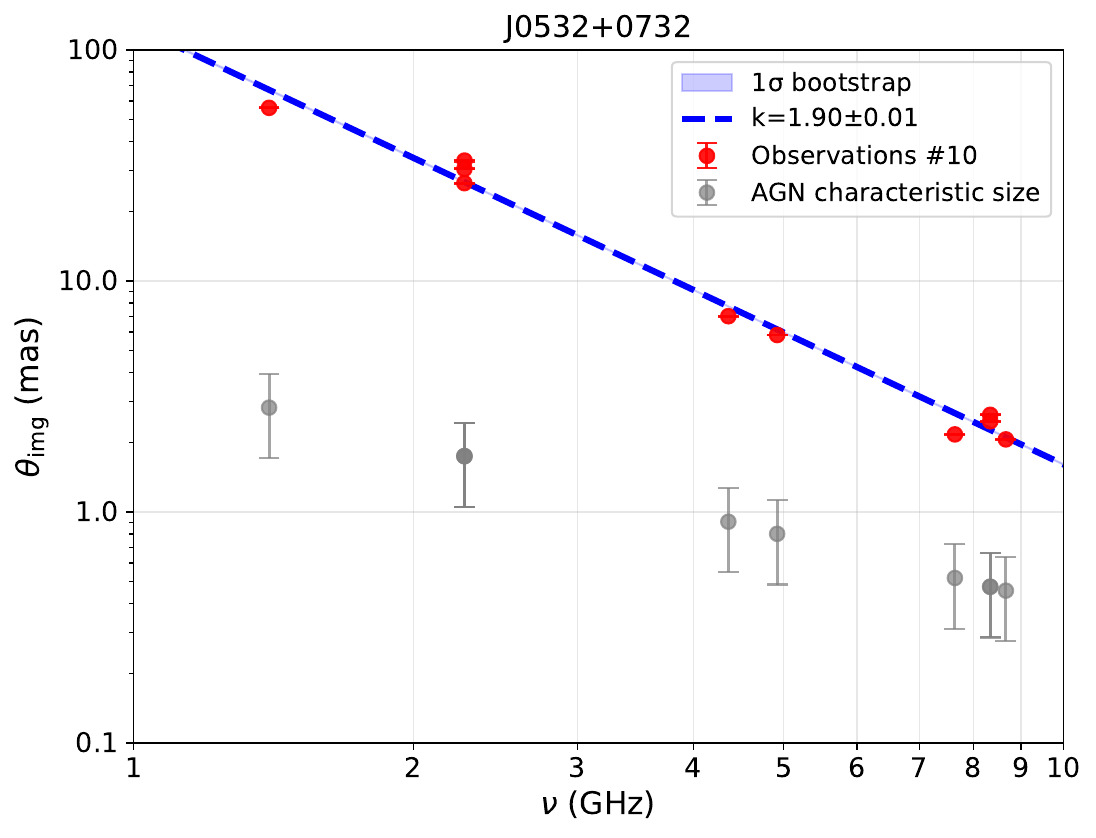}\vspace*{0cm}
    \caption{Frequency dependence of the angular size of the source J0532$+$0732. The gray shaded dots show estimates of the characteristic AGN angular size scaled to the observing frequency. This size is $3.9 \pm 1.6$~mas at $1$~GHz and was calculated as the median size of sources located above $15^\circ$ in absolute Galactic latitude.
} 
    \label{fig:freq_dependancy}
\end{figure}

To search for sources with signatures of scattering substructure, we restrict our analysis to those observed at least at two frequencies and not exceeding 8.7~GHz. We do not impose restrictions on Galactic latitude or longitude. A VLBI core is the compact, bright feature at the apparent origin of the jet, usually appearing as the dominant component in the AGN structure. In the absence of scattering, the angular size of the VLBI core is determined by its intrinsic structure and the opacity of the synchrotron emission. The frequency ($\nu$) dependence of the VLBI core size is given by $\theta_\mathrm{img}\propto\nu^{-k}$, where $k$ is the power-law index of the angular size frequency dependence, with $k\approx1$ for the intrinsic source size under the assumption of a conical jet shape \citep{BK79, Konigl1981}. However, when diffractive scattering dominates and causes angular broadening, the frequency dependence becomes steeper and follows a power law with $k\approx2$ \citep[e.g.,][]{Cordes86,Rickett1990,Armstrong95}. Thus, a value of $k$ close to 2 is a reliable marker of scattering and allows us to distinguish sources with a dominant ISM contribution from those dominated by their intrinsic jet structure. To estimate $k$, we used the obtained $\theta_\mathrm{img}$ values at all observational epochs without averaging. Figure~\ref{fig:freq_dependancy} illustrates the expected frequency dependence of the angular size for a source dominated by scattering, using J0532$+$0732 as an example.

Thus, the index $k$ was estimated for 8743 sources observed at least at two frequencies not exceeding 8.7~GHz. We include in the final sample of scattered sources only those for which the measured power-law index lies in the range $1.9\pm0.4$. These boundary values were chosen as a compromise between the need to reject sources with no significant scattering or outliers that likely arise from poorly constrained fits rather than extreme scattering. Thus, the sample was reduced to 1074 sources. It is worth noting that the majority of sources have only two-frequency observations, predominantly at 2 and 8~GHz. Therefore, indications of the presence of diffractive scattering are found for 12\% of the sources.

As mentioned above, the presence of a significant non-zero signal on long baselines can be interpreted as evidence of substructure arising from refractive scattering on interstellar medium inhomogeneities (e.g., JG15). We performed an automated search for sources with signatures of substructure contribution on long baselines. We used calibrated VLBI data from the Astrogeo database (typically processed in AIPS), to which we applied vector averaging of complex visibilities within each scan. Within a single scan, all measurements were grouped by baseline and frequency, after which weighted averages of the real and imaginary parts of the complex visibility were computed. The amplitude and phase were derived from the averaged components, and their uncertainties were estimated as the standard errors of the mean. This approach provides statistically robust estimates for each baseline in each scan.

To perform the automated search for sources with signatures of refractive scattering, we adopted the following assumptions:

\begin{enumerate}
\item \textbf{Dominance of scattering.} We assume that scattering is significant if the observed angular size follows the relation $\theta_\mathrm{img} \propto \nu^{-k}$, where $k$ is close to 2. We also assume that the source structure at low frequencies is well described by a single circular Gaussian component. If a single Gaussian component describes the data poorly (i.e., residual deviations significantly exceed the noise level), this may indicate that scattering is not significant enough and the observed structure is determined by intrinsic morphology. We excluded from further analysis those measurements for which a single Gaussian component provided a inadequate representation of the visibility data, typically due to the presence of resolved jet-like structures or significant residuals.

\item \textbf{Use of a priori AGN size information.} Equation~\ref{eq:JG19} contains two degenerate parameters: the observed angular size $\theta_{\rm img}$ and the scattering size $\theta_{\rm scatt}$. At the first selection stage, we estimated $\theta_{\rm img}$ as the median angular size over all observing epochs. To estimate $\theta_{\rm scatt}$, defined as $\theta_{\rm scatt} = \sqrt{\theta_{\rm img}^2 - \theta_{\rm int}^2}$, we use a priori information based on the statistics of AGN sizes for sources out of the Galactic plane (above $15^\circ$ in absolute Galactic latitude). The median angular size at 1~GHz is $\theta_{\rm int} = 3.9 \pm 1.6$~mas, based on the current version of the Astrogeo database available at the time of writing. The frequency dependence of the intrinsic size follows $\theta_{\rm int}(\nu) \propto \nu^{-1}$ for a conical jet \citep{BK79, Konigl1981}. Thus, using $\theta_{\rm img}$ and $\theta_{\rm int}$ for each source, as well as measured visibility amplitudes per baseline length, we can calculate the expected $\sigma_{\rm ref}(\mathbf{b})$ level within the JG15 framework.

\item \textbf{Use of all epochs.} We use all available observing epochs for each source and analyse the epoch-combined visibility function. The key assumption is that refractive substructure should be detectable over several years \citep{Plavin2026}. In contrast, the intrinsic structure of AGN typically shows significant variability on yearly timescales due to the internal evolution of relativistic jets \citep{Lister2021}. Thus, if the signal on long baselines is persistently present in observations from different epochs, this provides a strong argument for its refractive origin.
\end{enumerate}

The uncertainties of all parameters estimated in this work, including the power-law index of the frequency dependence $k$, Gaussian parameters, and others, were evaluated using the bootstrap method \citep{Efron1993}. The standard deviation of the resulting parameter distributions was adopted as their $1\sigma$ uncertainty. We also use letter designations for frequency bands throughout the paper, which generally correspond to the following mean observing frequencies: S (2.3~GHz), C (5.0~GHz), and X (8.6~GHz).

\section{Results and discussion}
\label{s:results}

\subsection{Detection of scattering substructure signatures}
\label{ss:substructure_detection}

\begin{figure*}
    \centering
    \includegraphics[width=0.7\linewidth]{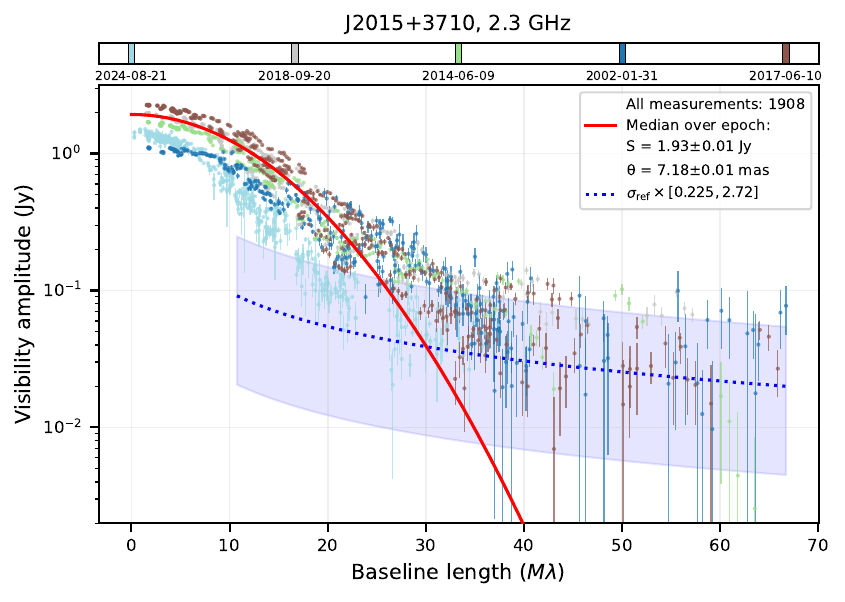}\vspace*{0cm}
    \caption{Visibility amplitude as a function of baseline projection for J2015$+$3710 (shown as an illustrative example). This figure demonstrates the search for sources with refractive scattering signatures using all available observation epochs at a given frequency. Points of the same colour represent observations from a single epoch, with the top panel listing all available epochs for this source in the Astrogeo database. The red line corresponds to the median angular size $\theta_{\rm img}$ of the source. The blue dashed line represents the mean predicted refractive noise level $\sigma_{\rm ref}(\mathbf{b})$, and the blue shaded region indicates the 95\% Rayleigh distribution interval defined as $[0.225, 2.72] \times \sigma_{\rm ref}(\mathbf{b})$. The predicted substructure band starting from the baseline where the Gaussian model first falls to 25\% of its peak.
} 
    \label{fig:radplot_J2015+3710}
\end{figure*}

\begin{figure}
    \centering
    \includegraphics[width=1\linewidth]{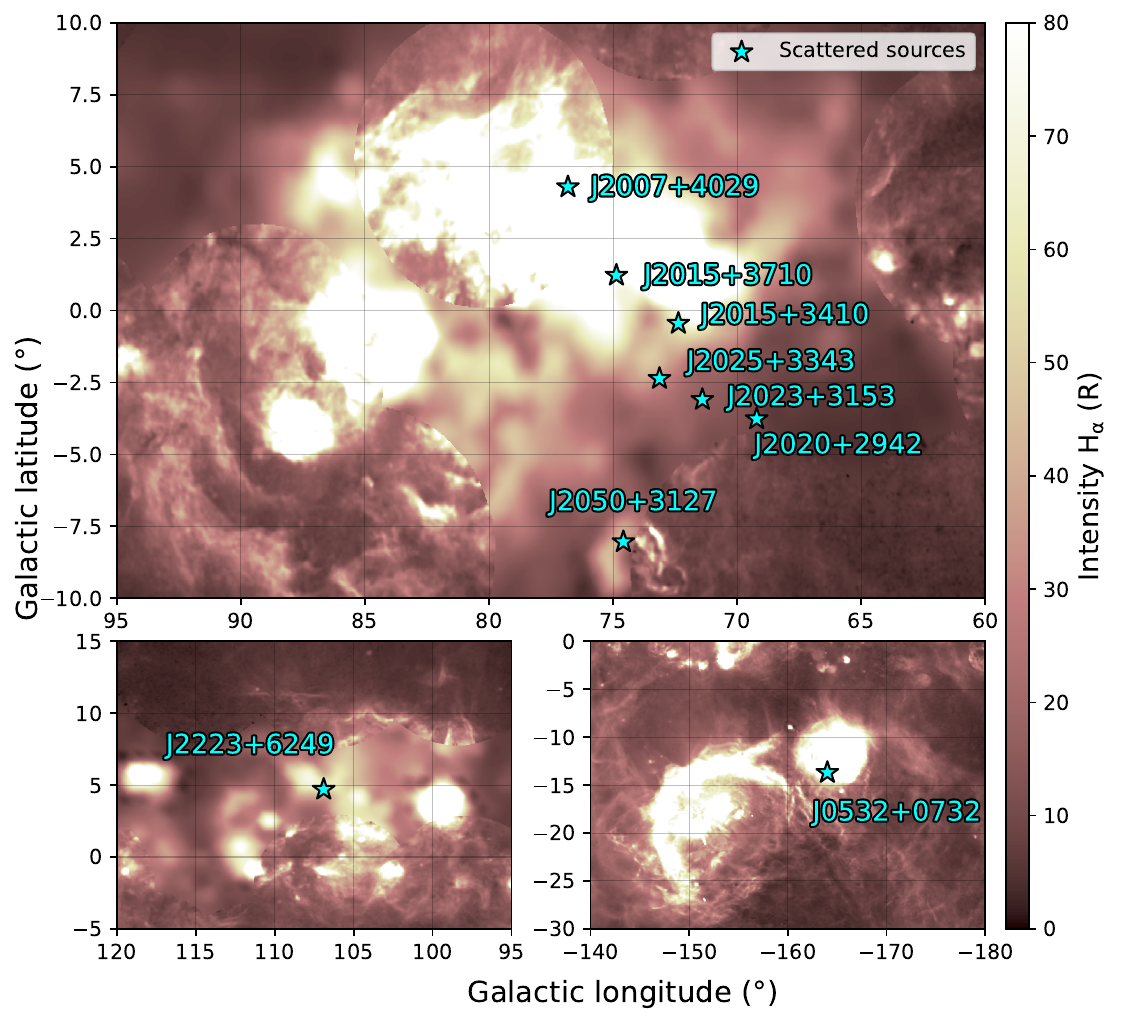}\vspace*{0cm}
    \caption{Sky distribution of scattered sources with refractive scattering substructure signatures. All detected sources are predominantly located in one region -- in the direction of the Cygnus constellation (upper panel). Only two sources, J2223$+$6249 and J0532$+$0732, are located in other directions (bottom panel). The colour map represents the $\mathrm{H_\alpha}$ intensity distribution in Rayleighs (1~R $= 10^6$~photons~cm$^{-2}$~s$^{-1}$~sr$^{-1}$).
} 
    \label{fig:sources}
\end{figure}

Taking into account all the assumptions mentioned in Section~\ref{s:sample} for $1074$ scattered sources in our sample, we compared the observed visibility amplitudes with the calculated $\sigma_{\rm ref}(\mathbf{b})$ level at frequencies ranging from $1.4$ to $8.7$~GHz. As a result of the systematic search, we selected sources for which the majority of data on long baselines fall within the 95\% interval $\sigma_{\rm ref}$. Figure~\ref{fig:radplot_J2015+3710} presents the observed visibility amplitude as a function of baseline projection for the quasar J2015$+$3710, illustrating the cases selected for further analysis. This figure shows that the majority of the data on baselines greater than $40\,M\lambda$ fall within the $[0.225, 2.72] \times \sigma_{\rm ref}(\mathbf{b})$ interval, and the effect persists over multiple epochs. We identified nine sources that satisfy these criteria. Notably, one of them is J2007$+$4029 -- the object from \citet{Plavin2026} in which scattering substructure was first detected with ground-based VLBI. Thus, the result of \citet{Plavin2026} is successfully reproduced in our analysis, confirming the validity of our approach. We do not analyse this source further, as it has been discussed in detail by \cite{Plavin2026}. Figure~\ref{fig:sources} shows the positions of these sources on the sky in Galactic coordinates. As we can see, the majority of sources are located in the direction of the Cygnus constellation, with only two AGNs outside this region. This once again highlights the particularly prominent scattering in this direction \citep{Fey1989, Gabani2006, Koryukova2023, Koryukova2025}. In addition to J2007$+$4029, our search also identified the quasar J2025$+$3343. For this source, VLBI data first revealed anisotropic refractive scattering with theoretically predicted multiple image formation \citep{Pushkarev2013}, as well as an extreme scattering event in Owens Valley Radio Observatory data at 15~GHz. Table~\ref{tab:measurements} summarizes the information on the nine detected sources: coordinates, epoch-averaged frequencies, flux densities, angular sizes and the estimated index $k$. The full set of individual measurements used to estimate the power-law index $k$ is provided in Appendix~\ref{s:appendix_sizes}. The background colour in Figure~\ref{fig:sources} shows the $\mathrm{H_\alpha}$ emission intensity map, tracing the distribution of hot ionized interstellar gas \citep{Finkbeiner03}. We used publicly available $\mathrm{H_\alpha}$ data\footnote{\url{https://faun.rc.fas.harvard.edu/dfink/skymaps/}} in Cartesian projection with 6~arcmin resolution.

\begin{table*}
\centering
\caption{Information on scattered sources: Galactic coordinates, epoch-averaged frequencies, flux densities and angular sizes defined as the full width at half maximum of fitted Gaussian component, and the estimated power-law index $k$ of the angular size frequency dependence.}
\small
\setlength{\tabcolsep}{2pt}
\begin{tabular}{lccccccccccccc}
\hline
J2000 name   & $l$   & $b$    & $\nu$ & $S$ & $\theta$ & $\nu$ & $S$ & $\theta$ & $\nu$ & $S$ & $\theta$ & $k$ \\
 & ($^\circ$) & ($^\circ$) & (GHz) & (mJy) & (mas) & (GHz) & (mJy) & (mas) & (GHz) & (mJy) & (mas) &  \\
\hline
J0532$+$0732  & 196.8 & $-13.7$&2.3& $2040.7 \pm 17.9$& $30.46 \pm 0.11$ &4.6& $1328.1 \pm 4.5$ & $6.42 \pm 0.02$ &8.2& $1280.7 \pm 2.3$ & $2.31 \pm 0.01$ & $1.90 \pm 0.01$ \\
J2007$+$4029$^a$& 76.8  & 4.3    &2.3& $2302.3 \pm 4.5$ & $12.98 \pm 0.02$ &\ldots & \ldots       & \ldots          &8.1& $1829.8 \pm 1.4$ & $1.45 \pm 0.01$ & $1.69 \pm 0.01$ \\
J2015$+$3410  & 72.4  & $-0.4$ &2.3& $513.0 \pm 1.7$  & $9.66 \pm 0.04$  &\ldots & \ldots       & \ldots          &8.6& $353.6 \pm 1.3$  & $1.23 \pm 0.01$ & $1.55 \pm 0.01$ \\
J2015$+$3710  & 74.9  & 1.2    &2.3& $1929.0 \pm 2.0$ & $7.19 \pm 0.01$  &4.4& $2758.6 \pm 3.9$ & $2.17 \pm 0.01$ &8.4& $2348.6 \pm 1.1$ & $0.56 \pm 0.01$ & $2.09 \pm 0.01$ \\
J2020$+$2942$^b$ & 69.2  & $-3.8$ &2.3& $684.3 \pm 1.0$  & $8.04 \pm 0.02$  &4.4& $405.1 \pm 1.6$  & $9.61 \pm 0.03$ &8.5& $107.2 \pm 1.2$  & $0.39 \pm 0.05$ & $2.19 \pm 0.02$ \\
J2023$+$3153  & 71.4  & $-3.1$ &2.3& $1423.4 \pm 1.7$ & $4.35 \pm 0.01$  &4.4& $1219.7 \pm 2.4$ & $1.52 \pm 0.01$ &8.4& $981.5 \pm 1.0$  & $0.71 \pm 0.01$ & $1.55 \pm 0.01$ \\
J2025$+$3343  & 73.1  & $-2.4$ &2.3& $1003.3 \pm 2.8$ & $10.60 \pm 0.04$ &4.4& $1676.0 \pm 4.3$ & $4.15 \pm 0.01$ &8.4& $2091.1 \pm 1.5$ & $1.06 \pm 0.01$ & $1.73 \pm 0.01$ \\
J2050$+$3127  & 74.6  & $-8.1$ &2.3& $589.0 \pm 2.5$  & $8.61 \pm 0.05$  &4.4& $483.1 \pm 1.8$  & $3.24 \pm 0.01$ &8.4& $425.8 \pm 0.8$  & $0.92 \pm 0.01$ & $1.69 \pm 0.01$ \\
J2223$+$6249  & 106.9 & 4.7    &2.3& $182.9 \pm 2.4$  & $9.74 \pm 0.24$  &4.4& $218.9 \pm 0.7$  & $3.08 \pm 0.01$ &8.3& $202.3 \pm 0.7$  & $0.98 \pm 0.01$ & $1.84 \pm 0.02$ \\
\hline
\end{tabular}
\begin{tablenotes}
    \item Note: $S$ and $\theta$ are given at the frequency listed in the preceding $\nu$ column.
    \item $a$ -- substructure detected in \citet{Plavin2026}.
    \item $b$ -- J2020$+$2942 is also listed as J2020$+$294A in the Astrogeo database.
\end{tablenotes}

\label{tab:measurements}
\end{table*}

For each of the eight selected sources, we performed a thorough visual inspection of the observed visibility functions at all available epochs and frequencies (L, S, C-bands). For each epoch and frequency, we estimated the angular size $\theta_{\rm img}$ and used it to calculate the predicted refractive noise level $\sigma_{\rm ref}(\mathbf{b})$. This analysis showed that for all candidate sources, a non-zero signal on long baselines is present at least two epochs and at two observing frequencies (in various combinations of L, S, C-bands). This is an important argument in favour of the physical nature of the observed phenomenon, as opposed to a random fluctuation or processing artifact, since its presence does not depend on a specific epoch or frequency.

\begin{figure*}
    \centering
    \includegraphics[width=0.44\linewidth]{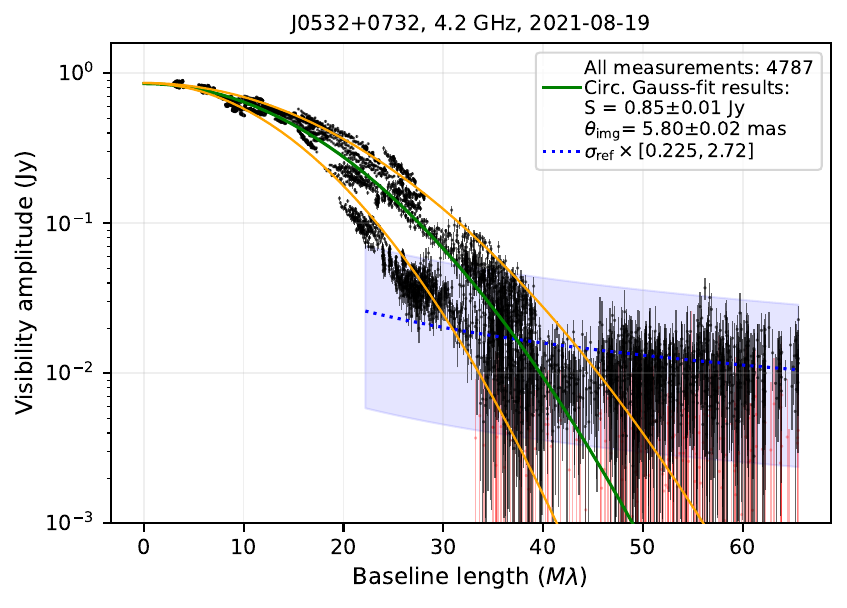}\vspace*{0cm}
    \includegraphics[width=0.44\linewidth]{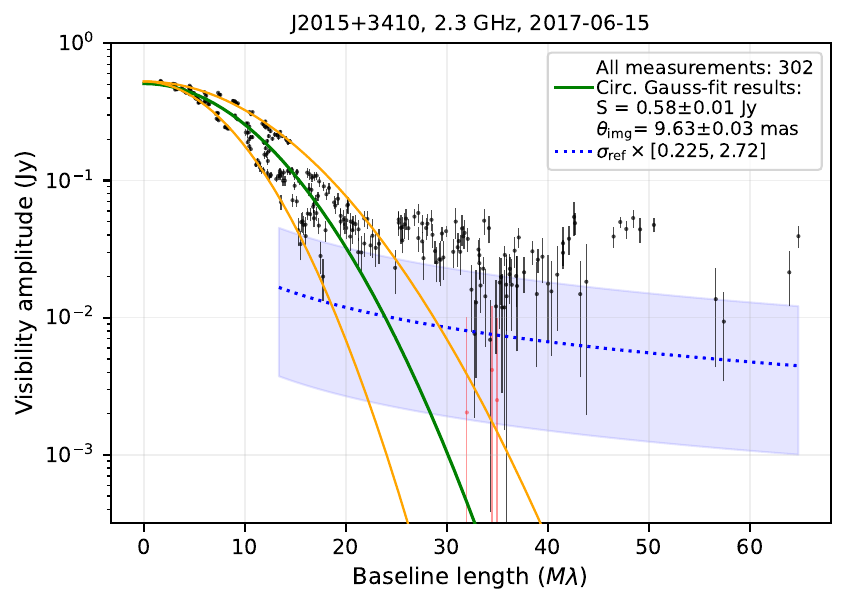}\vspace*{0cm}
    \includegraphics[width=0.44\linewidth]{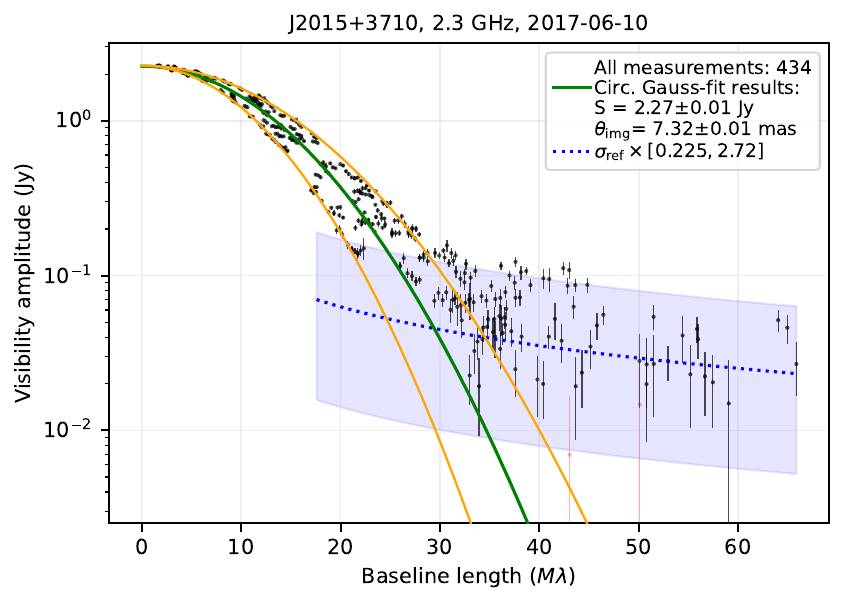}\vspace*{0cm}
    \includegraphics[width=0.44\linewidth]{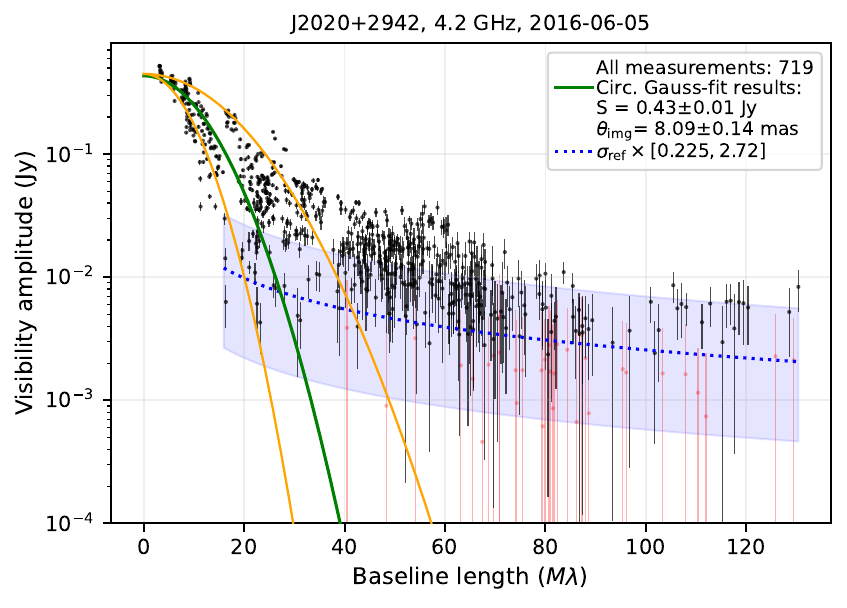}\vspace*{0cm}    
    \includegraphics[width=0.44\linewidth]{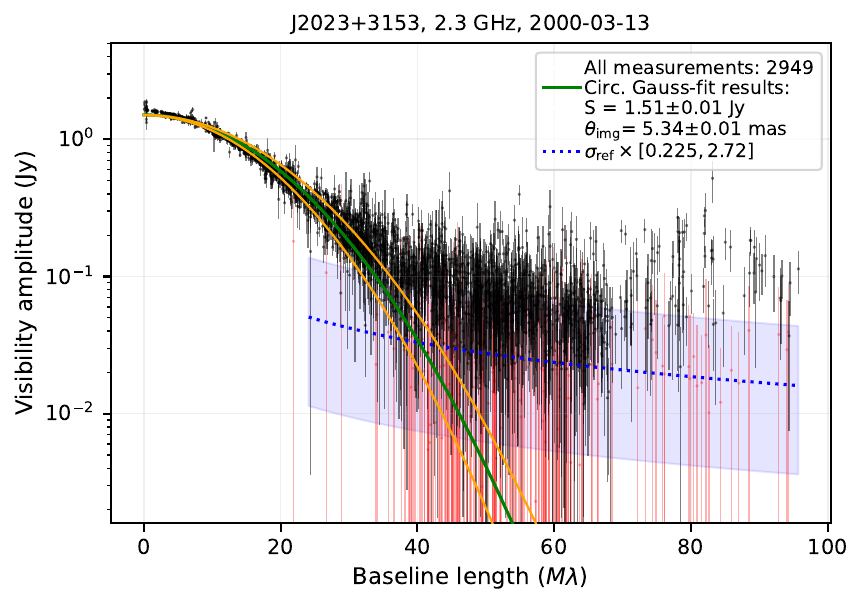}\vspace*{0cm} 
    \includegraphics[width=0.44\linewidth]{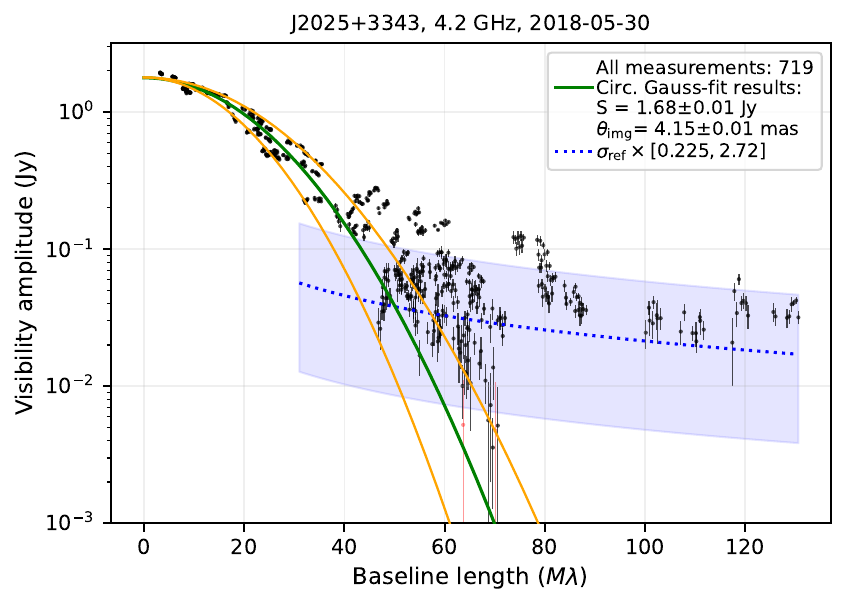}\vspace*{0cm}
    \includegraphics[width=0.44\linewidth]{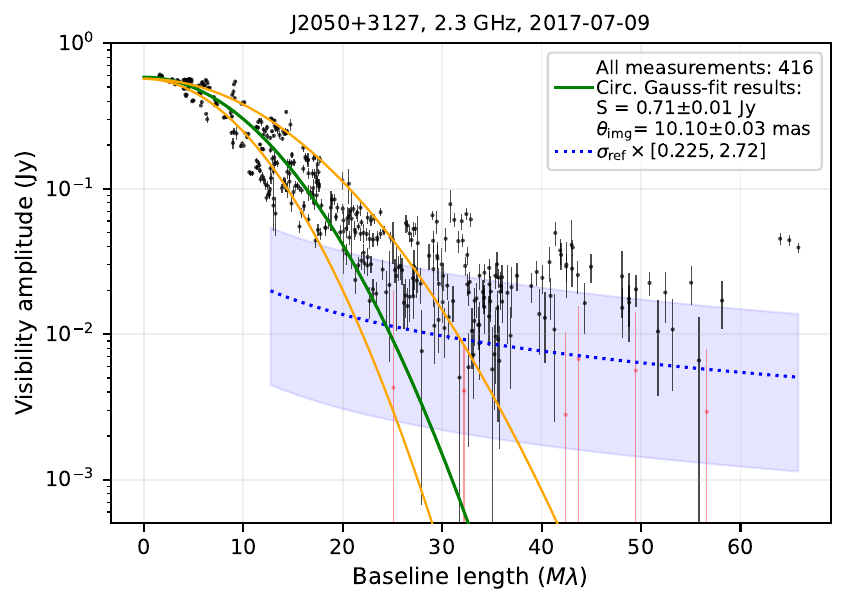}\vspace*{0cm}
    \includegraphics[width=0.44\linewidth]{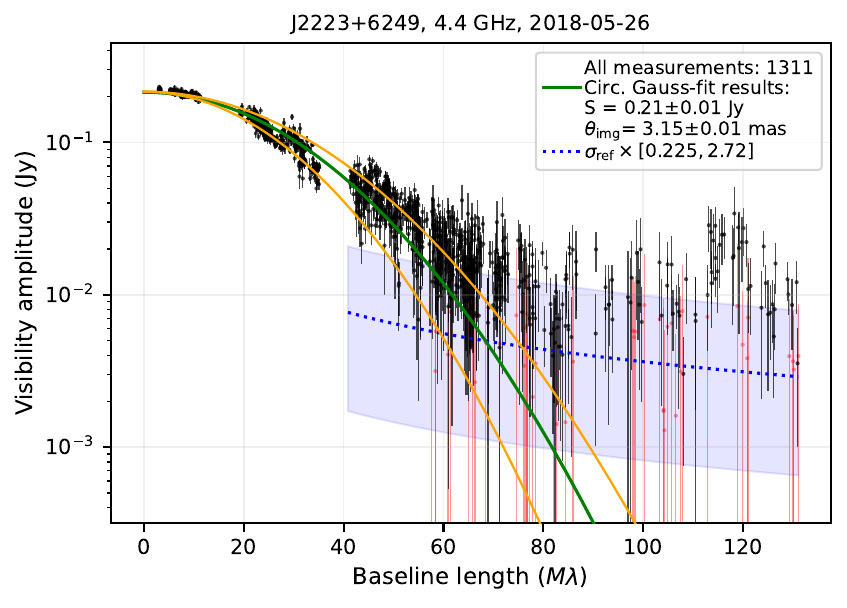}\vspace*{0cm}

    \caption{Visibility amplitude as a function of baseline length projection. Black points show scan-averaged observations from a single epoch where the effect is the most significant. The green line shows the estimated angular size obtained from a single circular Gaussian component fit (orange curves show elliptical Gaussian fits). The blue dashed line indicates the mean predicted refractive noise level $\sigma_{\rm ref}(\mathbf{b})$, and the blue shaded region represents the 95\% Rayleigh distribution interval $[0.225, 2.72] \times \sigma_{\rm ref}(\mathbf{b})$. The observing source, frequency and epoch are indicated in each panel title. Red points show measurements with errors exceeding 100\%.
} 
    \label{fig:radplots}
\end{figure*}

Following the visual inspection, we selected the best observing epoch for each source, defined as the epoch with the greatest number of measurements consistent with the $[0.225, 2.72] \times \sigma_{\rm ref}(\mathbf{b})$ interval. Figure~\ref{fig:radplots} presents the visibility amplitude as a function of baseline length for eight sources and selected epochs in S and C-bands. The green line shows the fitted Gaussian distribution; this $\theta_{\rm img}$ measurement was used to calculate the predicted refractive noise level $\sigma_\mathrm{ref}(\mathbf{b})$. Red points indicate measurements with errors exceeding 100\% of measured amplitude.

The presence of data points above the predicted $95\%$ interval of the JG15 model (e.g., for J2015$+$3410, J2023$+$3153, or J2223$+$6249 in Figure~\ref{fig:radplots}) may have two main explanations. First, the model assumes a specific set of parameters for the scattering screen (e.g., Kolmogorov turbulence with a fixed inner scale and spectral index), which may not accurately describe the actual conditions along a given line of sight. Second, this may indicate that the intrinsic source structure contributes significantly at this frequency. To distinguish between these possibilities, multi-epoch observations with high sensitivity are crucial, since refractive substructure is expected to remain stable over timescales of years, while intrinsic jet components typically exhibit variability. A detailed investigation of the source structure is beyond the scope of this work; nevertheless, even with such information, the two effects would remain partially degenerate, making it difficult to unambiguously determine whether scattering substructure is present or not.

\subsection{Long-baseline detections}
\label{ss:long_base_detections}

\begin{figure*}
    \centering
    \includegraphics[width=0.49\linewidth]{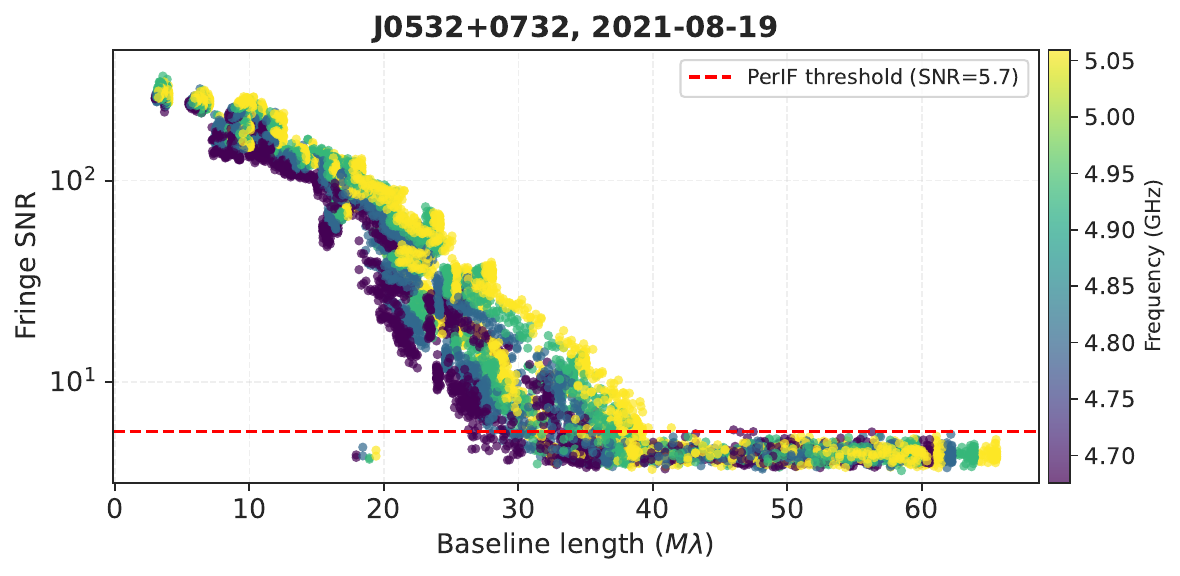}\vspace*{0cm}
    \includegraphics[width=0.49\linewidth]{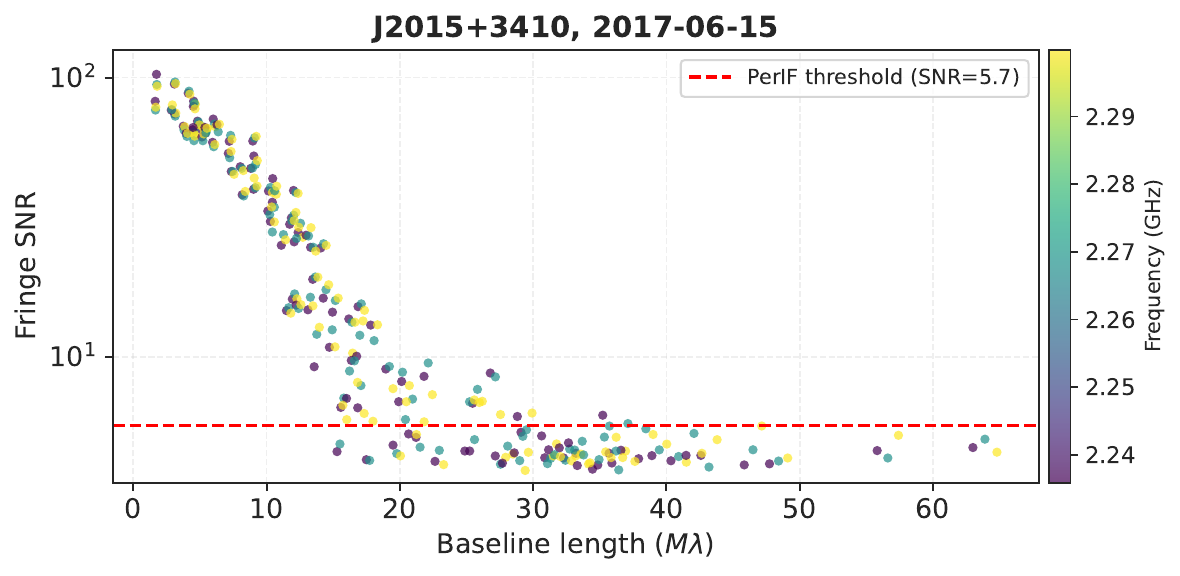}\vspace*{0cm}
    \includegraphics[width=0.49\linewidth]{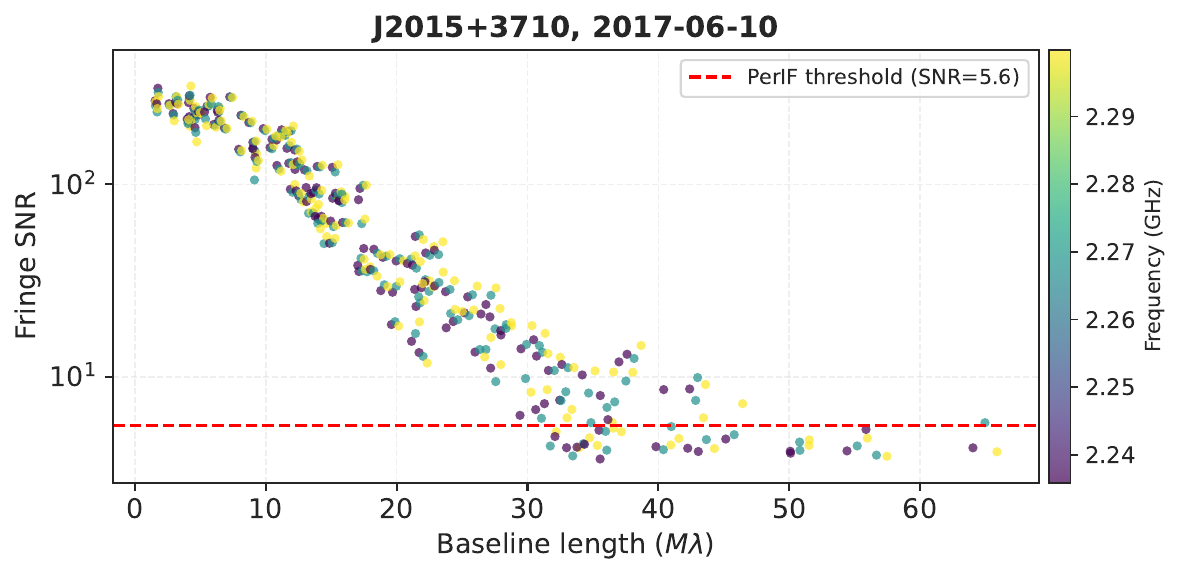}\vspace*{0cm}
    \includegraphics[width=0.49\linewidth]{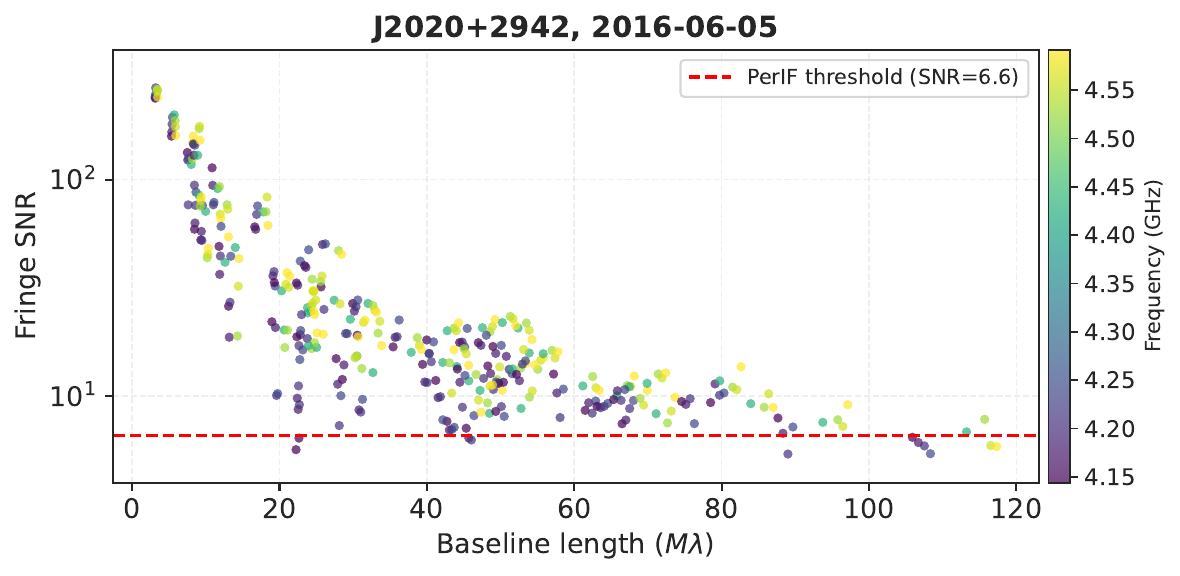}\vspace*{0cm}
    \includegraphics[width=0.49\linewidth]{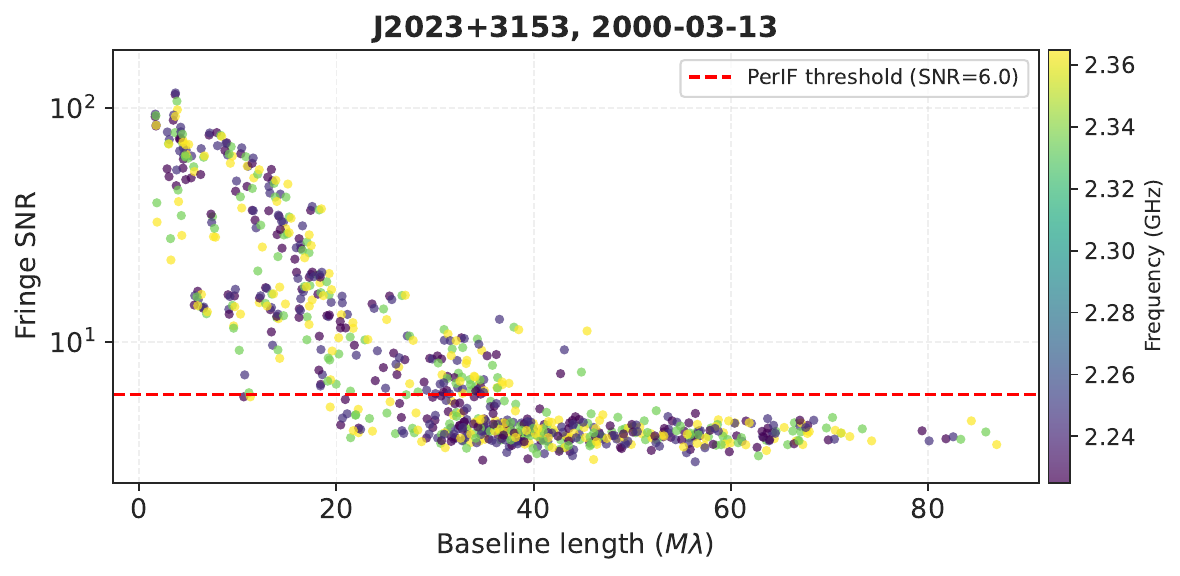}\vspace*{0cm}
    \includegraphics[width=0.49\linewidth]{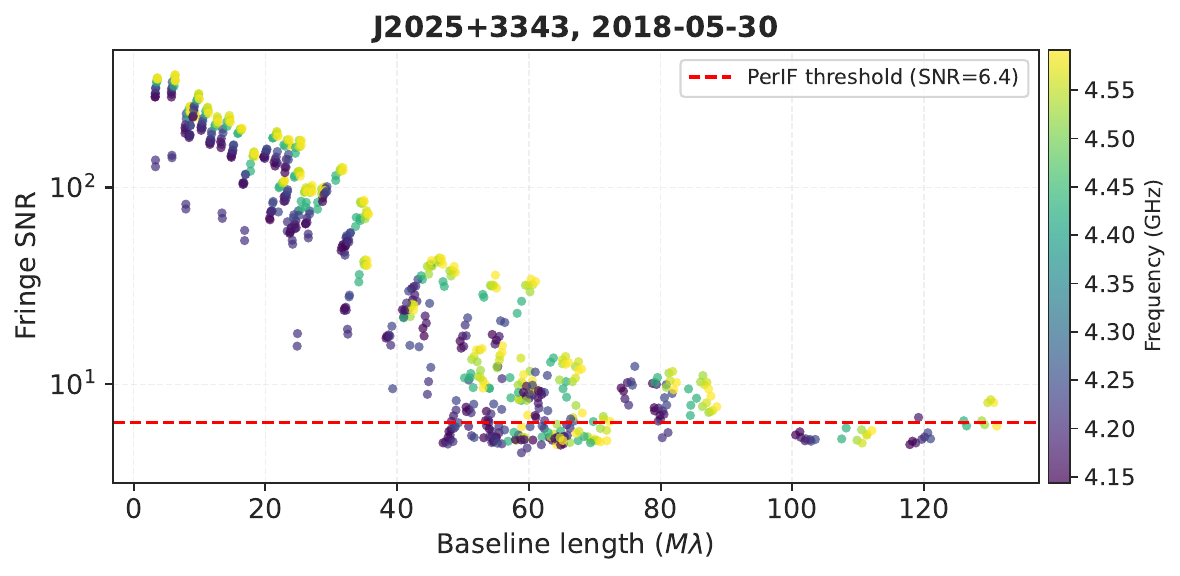}\vspace*{0cm}
    \includegraphics[width=0.49\linewidth]{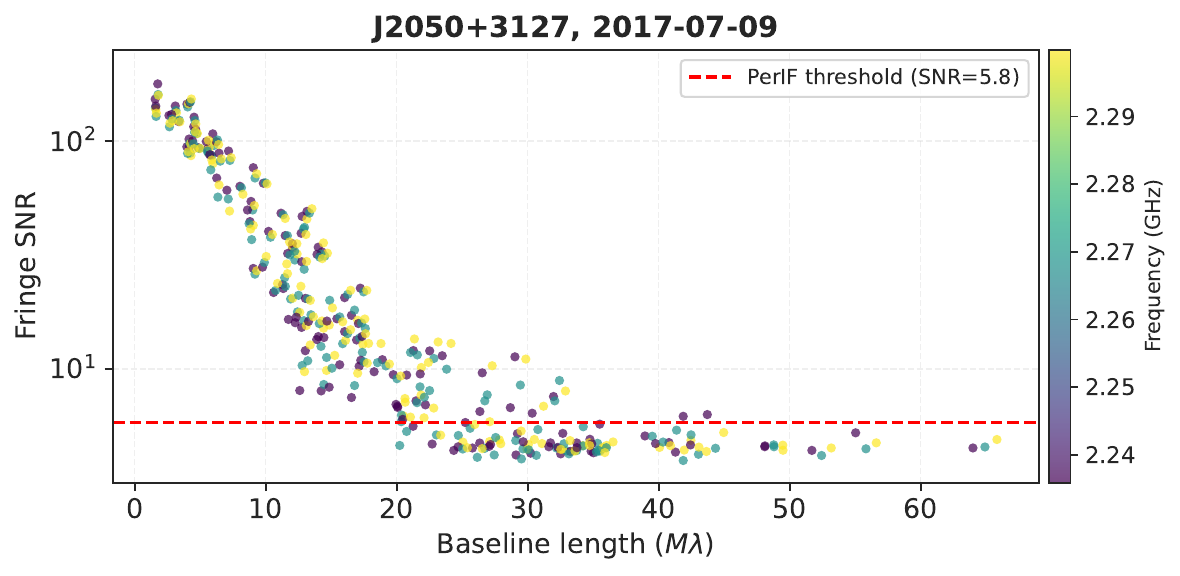}\vspace*{0cm}
    \includegraphics[width=0.49\linewidth]{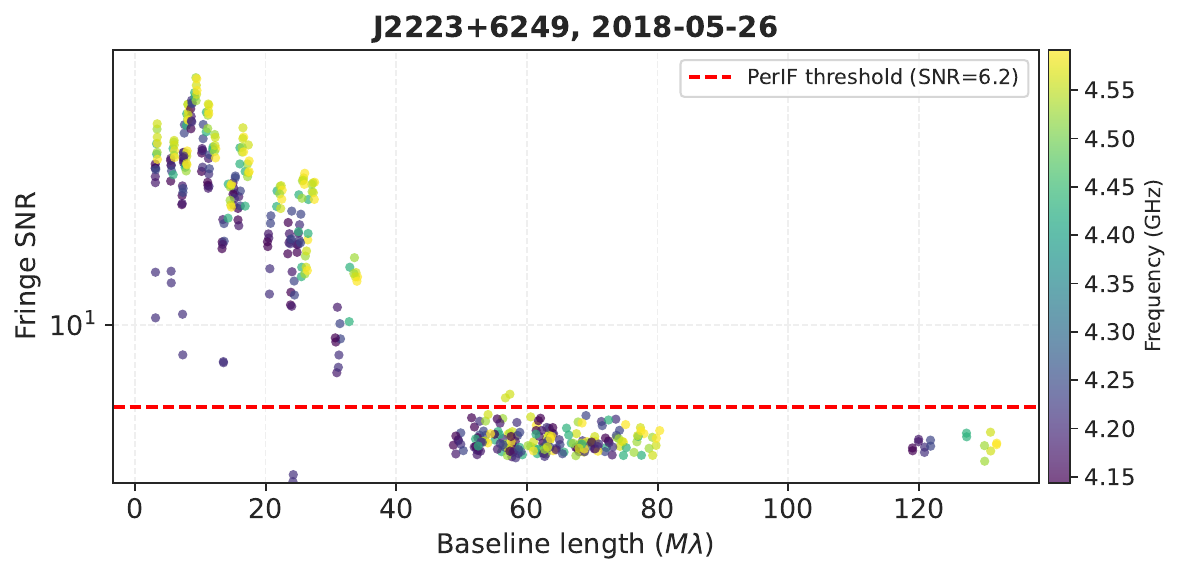}\vspace*{0cm}
    
    \caption{Fringe SNR as a function of baseline projection for scattered sources observed with ground-based VLBI. Point colours correspond to the observing frequency according to the colorbar on the right. The red line indicates the SNR above which detections are reliable according to our false-detection probability criterion for the \textit{per-IF} search. The observing epoch and frequency correspond to those listed for each source in Figure~\ref{fig:radplots}.
} 
    \label{fig:fft_results}
\end{figure*}

To verify the reliability of the long-baseline measurements, we applied the fringe search described in Section~\ref{ss:detections} to the raw correlator data of the selected sources and their best observing epochs. The results of the fringe search in \textit{per-IF} and \textit{combined} modes for eight candidate sources are presented in Table~\ref{tab:fft_results}. This table shows the fraction of detections for all baselines, as well as specifically for long baselines, defined as signals detected on baselines where the diffractive scattering Gaussian amplitude falls below 1/1000 of its peak value. The results show that two sources, J2025$+$3343 and J2020$+$2942, have a substantial fraction of long-baseline detections, reaching 100\% in \textit{combined} mode for both. This indicates that for J2025$+$3343 and J2020$+$2942, the presence of refractive substructure is reliably detected at the chosen epochs. In addition to these two sources, J2050$+$3127, J2015$+$3710, and J2015$+$3410 show significantly lower but non-zero fractions of long-baseline detections. For the other sources (J0532$+$0732, J2023$+$3153 and J2223$+$6249), the number of detections remains too low (or even zero) even after combining IFs, and cannot be considered reliable, being most likely due to statistical fluctuations. Thus, for five sources (J2025$+$3343, J2020$+$2942, J2050$+$3127, J2015$+$3710, and J2015$+$3410) the best available epochs provide compelling evidence of long-baseline detections, which may indicate the presence of scattering substructure in these objects. Figure~\ref{fig:fft_results} shows the fringe SNR as a function of baseline projection of scattered sources observed with ground-based VLBI. The panels show observations in S or C-bands at one of the available epochs. The red line indicates the SNR above which detections are reliable according to our false-detection probability criterion for the \textit{per-IF} search.

\begin{table*}
\centering
\small
\caption{Results of the baseline-based fringe search applied to the raw VLBA correlator data of the scattered sources.}
\begin{tabular}{lcccccrrrrr}
\hline
J2000 name & Exp. code & Epoch & Band   & Bandwidth & Scan dur. & \multicolumn{2}{c}{Total detections} & \multicolumn{2}{c}{Long baseline detections} \\ 
           &           &       &        & (MHz)  & (s)         & Per IF & Combined IF & Per IF & Combined IF \\
    (1)    &    (2)    &   (3) &   (4)  & (5)    &  (6)        &   (7)  & (8)         & (9)    & (10) \\
            
\hline
J0532$+$0732  & BZ087B  & 2021-08-19 & C & 128 & 20  & 7038/9632 (73\%) & 1899/2408 (79\%)   & 0/1126  (0\%)  & 2/278 (1\%)\\
J2015$+$3410  & UF001L  & 2017-06-15 & S & 32  & 98  & 238/408   (58\%) & 78/102    (76\%)   & 2/85    (2\%)  & 4/21  (19\%)\\
J2015$+$3710  & UF001K  & 2017-06-10 & S & 32  & 56  & 433/540   (80\%) & 128/135   (95\%)   & 9/45    (20\%) & 6/11  (55\%)\\
J2020$+$2942  & BP192G  & 2016-06-05 & C & 32  & 199 & 349/360   (97\%) & 45/45     (100\%)  & 174/183 (95\%) & 23/23 (100\%)\\
J2023$+$3153  & RDV20   & 2000-03-13 & S & 8   & 350 & 366/784   (47\%) & 102/196   (52\%)   & 0/112   (0\%)  & 0/26  (0\%)\\
J2025$+$3343  & BS262A  & 2018-05-30 & C & 32  & 50  & 618/720   (86\%) & 88/90     (98\%)   & 64/100  (63\%) & 10/10 (100\%)\\
J2050$+$3127  & UF001M  & 2017-07-09 & S & 32  & 147 & 328/540   (61\%) & 104/135   (77\%)   & 4/99    (4\%)  & 6/26  (23\%)\\
J2223$+$6249  & SB072A  & 2018-05-26 & C & 32  & 44  & 238/448   (53\%) & 35/56     (62\%)   & 0/16    (0\%)  & 0/2   (0\%)\\
\hline
\end{tabular}
\vspace{2mm}
\parbox{\textwidth}{\footnotesize
Column description: (1) J2000 source name; (2) experiment code; (3) observation epoch; (4) frequency band; (5) total bandwidth; (6) scan duration; (7),(8) total detections in \textit{per-IF} and \textit{combined} modes; (9),(10) detections on long baselines in \textit{per-IF} and \textit{combined} modes.\\
Note: We define long-baseline detections as signals detected on baselines where the diffractive scattering Gaussian falls below 1/1000 of its peak amplitude.
}
\label{tab:fft_results}
\end{table*}

Non-zero long-baseline amplitudes in the calibrated data are expected even where our fringe search finds no signal: standard processing retains these visibilities without a per-baseline detection test, so their amplitudes can be entirely noise-driven, and self-calibration can additionally bias them towards the source model. The magnitude of these effects varies across the Astrogeo database, which combines observations processed by different researchers with different software packages (AIPS, PIMA, etc.).

Despite the fact that the conservative baseline-based analysis did not reveal detections for all sources across all available epochs and frequencies, we consider these eight sources to be the most promising AGNs for targeted observations aimed at detecting and investigating refractive scattering substructure. These objects are priority targets for our future observations, the main goal of which will be to constrain the parameters of the scattering screen, including $r_{\rm in}$ and $\alpha$.

\subsubsection{Flux density of scattered sources}
\label{sss:flux_density}

\begin{figure*}
    \centering
    \includegraphics[width=0.49\linewidth]{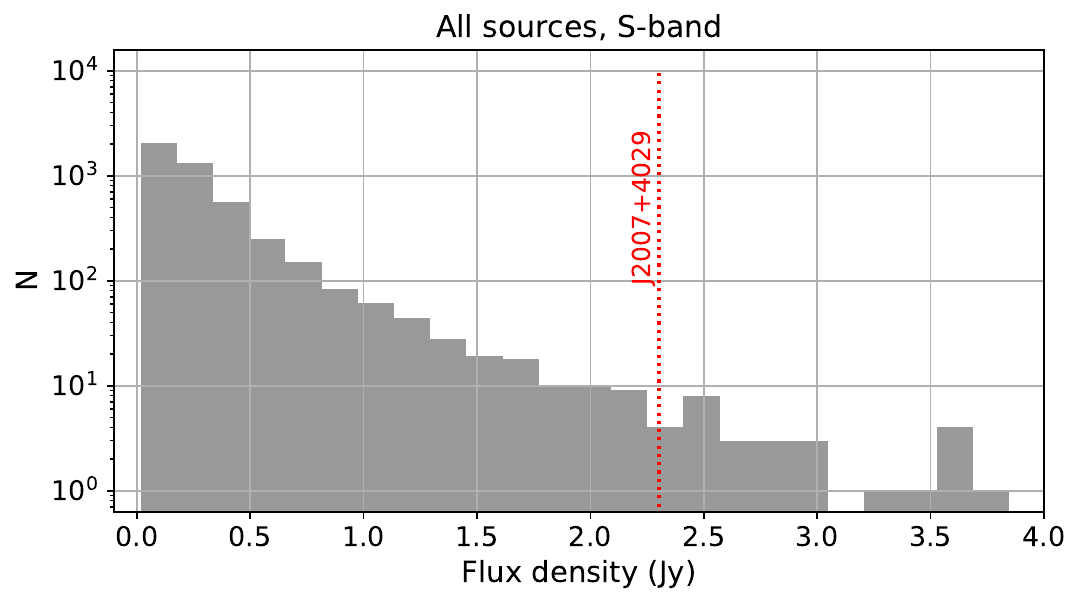}\vspace*{0cm}
    \includegraphics[width=0.49\linewidth]{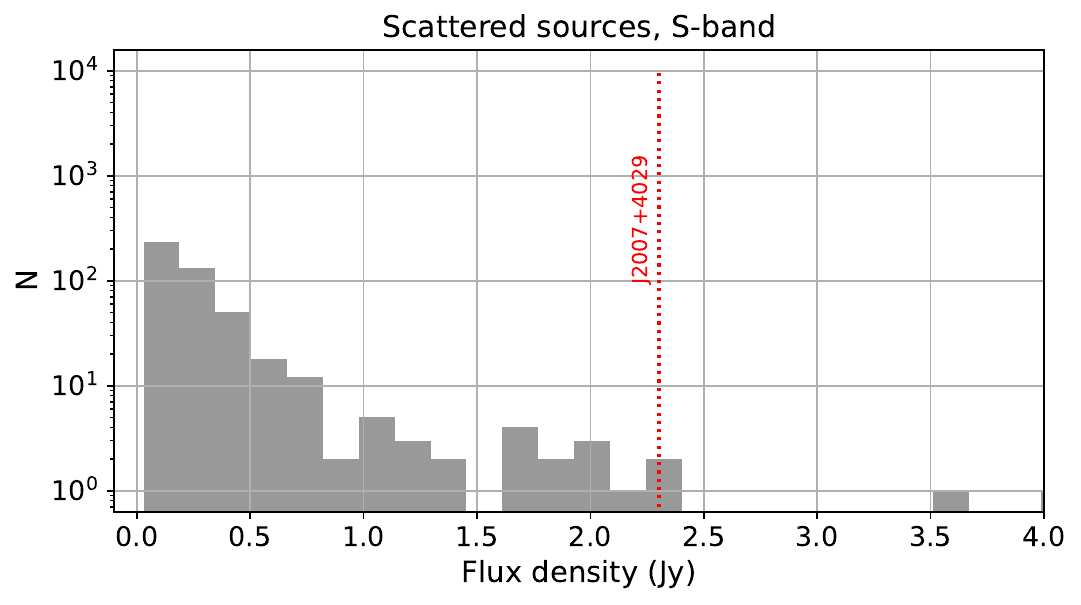}\vspace*{0cm}
    \includegraphics[width=0.49\linewidth]{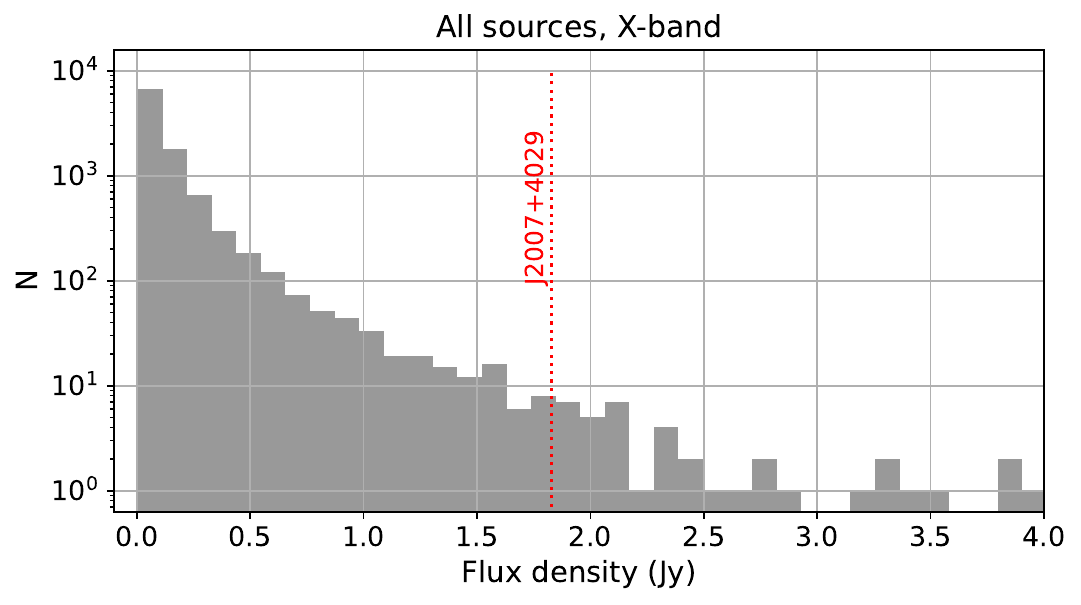}\vspace*{0cm}
    \includegraphics[width=0.49\linewidth]{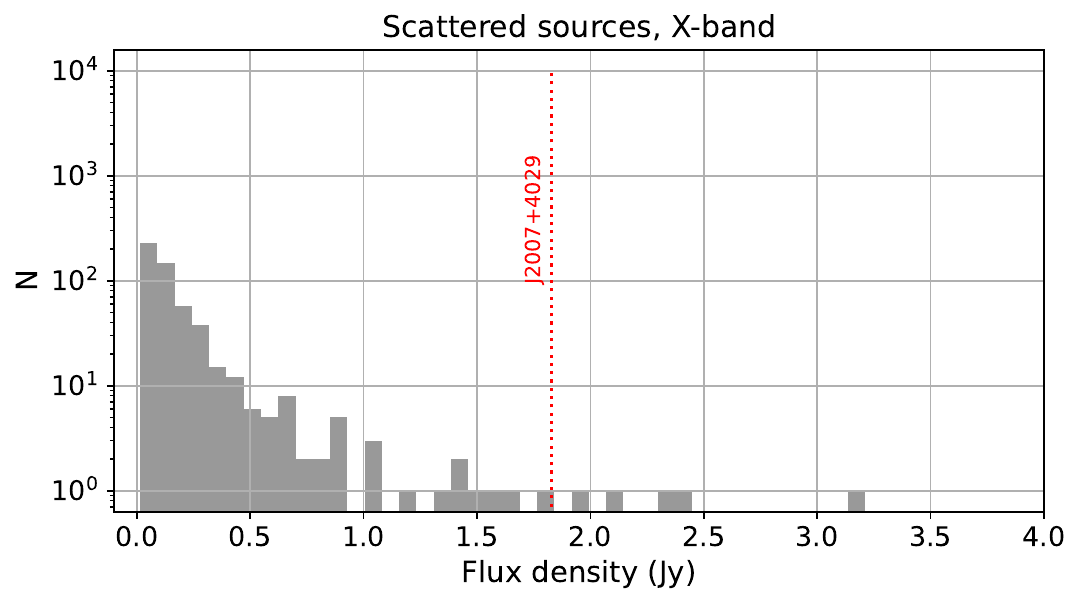}\vspace*{0cm}    
    \caption{Distributions of median over epochs flux densities. The left column shows the distributions for all sources in the Astrogeo database, while the right column shows only the 1074 AGNs selected as sources with diffractive scattering signatures. Data are presented for two frequency bands: S (2.3~GHz) and X (8.6~GHz). The red line marks the flux density value for J2007$+$4029. The vertical axis shows the number of sources $N$.
} 
    \label{fig:flux_densities}
\end{figure*}

While it is known that J2007$+$4029 has a high core flux density of about $2302$~mJy, the flux densities of the remaining eight sources in our sample are generally lower (see Table~\ref{tab:measurements}). Figure~\ref{fig:flux_densities} shows the distributions of median over epochs flux densities for the entire sample (left column) and for the 1074 sources with diffractive scattering signatures (right column). This figure shows that sources with a core flux density at low frequencies comparable to J2007$+$4029 are quite rare. The brightest sources in terms of flux density in our sample are J0532$+$0732 ($2040.7 \pm 17.9$~mJy) and J2015$+$3710 ($1929.0 \pm 2.0$~mJy). This result highlights why J2007$+$4029 is a unique object: its high brightness combined with strong scattering makes the refractive signal detectable even with ground-based VLBI.

Although our eight sources exhibit a steep frequency dependence of their sizes, and the measured visibility amplitudes on long baselines fall within the predicted interval of the JG15 model, lower brightness and the limited integration time (snapshot observing mode) prevents us from accumulating sufficient SNR for a reliable detection of the refractive substructure. Therefore, a key strategy for future observations is to increase the scan length, especially at lower frequencies (L and S-bands).

\subsection{Future prospects}
\label{ss:prospects}

Future observations of refractive substructure in a large sample of AGNs are promising for two main reasons. First, they will help recover the intrinsic jet structure and improve brightness temperature estimates, which are essential for understanding the physics of relativistic outflows \citep[e.g.,][]{Kovalev05, Lobanov2015}. Second, refractive scattering is a unique probe of the ionized interstellar medium along all lines of sight across the sky.

There are several strategies to achieve higher sensitivity in future VLBI observations for detecting and investigating refractive substructure:

\begin{enumerate}
\item \textbf{Improved baseline sensitivity.} The detection of refractive substructure in a larger sample of sources will ultimately depend on improving the baseline sensitivity, which is determined by individual telescopes, as well as the total bandwidth and the scan length. The current snapshot observations used in this work were limited by relatively short scans and narrow bandwidths, preventing us from achieving sufficient SNR for robust detection. Future instruments, such as the Next Generation Very Large Array (ngVLA, \citealt{Selina2018}) and Square Kilometre Array (SKA, e.g., \citealt{Kovalev2026}), will offer dramatic improvements in sensitivity and wide-bandwidth capabilities.

\item \textbf{Improved $uv$-coverage.} Increasing the number of antennas in an interferometric array significantly improves $uv$-coverage, since the number of independent baseline projections grows as $M(M-1)/2$, where $M$~is the number of antennas. This not only improves image reconstruction quality but also increases sensitivity to weak signals due to a larger number of independent measurements. Of particular importance for our study is the availability of short baselines, since they provide recovery of the extended structure of scattered sources and accurate determination of their angular sizes. In this regard, observations using the European VLBI Network (EVN) in combination with Multi-Element Radio Linked Interferometer Network (MERLIN) are especially promising, as they already allow correlation of data from all antennas simultaneously, providing significantly denser $uv$-coverage compared to the VLBA array (up to 10 antennas), on whose data our work is based. In the longer term, facilities like VLBA combined with ngVLA will offer even greater opportunities for such studies, offering a large number of antennas and unprecedented sensitivity. Improved coverage is not only beneficial for sensitivity, but is also critical for studying the anisotropic nature of the scattering screen, which carries information about the direction of the Galactic magnetic field and the properties of MHD turbulence.

\end{enumerate}

\section{Conclusion}
\label{s:conclusion}

In this work, we have performed the first systematic search for refractive scattering substructure in active galactic nuclei using ground-based VLBI. Based on the analysis of more than 19\,000 sources from the Astrogeo database, we selected 1074 objects with a steep frequency dependence of angular size, indicating signatures of diffractive scattering. Comparison of the observed visibility amplitudes on long baselines with the predictions of the JG15 model revealed eight candidates for which a significant fraction of the data are consistent with the expected level of refractive noise. We found that evidence for scattering substructure in these sources is present at least for two epochs and at two observing frequencies, which argues against a spurious origin of the effect.

Baseline-based fringe search revealed reliable long-baseline detections for two sources, J2020$+$2942 and J2025$+$3343, as well as partial detections for J2015$+$3410, J2015$+$3710, and J2050$+$3127, providing clear evidence for the presence of scattering substructure on long baselines. For the remaining three sources (J0532$+$0732, J2023$+$3153 and J2223$+$6249), however, no such signal is detected, which we attribute primarily to the limited sensitivity of the available data. Nevertheless, all eight candidates remain promising targets for future observations with improved instrumental sensitivity.

The higher sensitivity is achievable through a combination of strategies (in particular, observations with the EVN+MERLIN or VLBA+ngVLA programs in a wide frequency band and with long integration times), may be sufficient to move from the status of ``candidates'' to reliable detections all of eight sources. These objects are priority targets for our future VLBI observations aimed at the systematic study of refractive scattering in AGN through the Galactic interstellar medium.

Remarkably, despite the systematic search across the entire sky, most of the candidates are located in a relatively small region -- in the direction of the Cygnus constellation. This fact once again indicates the presence of turbulent plasma in this direction. The sources J2223$+$6249 and J0532$+$0732 are particularly interesting for future studies, since the lines of sight toward them lie outside the Cygnus region and therefore probe different, less explored parts of the interstellar medium. For all of the sources, multi-frequency observations with sufficiently high sensitivity are needed, which will allow reliable detection of refractive substructure and estimating the parameters of the scattering screens along these lines of sight, including the inner turbulence scale ($r_\mathrm{in}$) and the power-law index of turbulence scale spectrum ($\alpha$). Multi-epoch observations will also allow studying the temporal variability of these parameters.

Our work demonstrates that the systematic study of refractive substructure in AGN is approaching a new phase, where carefully designed high-sensitivity VLBI observations of sources of particular interest become the key to further progress. In the longer term, using AGNs as cosmic probes of the ISM will complement traditional methods of studying interstellar turbulence through pulsar observations \citep[e.g.,][]{Rickett1990, Elmegreen_2004, Scalo2004} and will enable the construction of a more accurate distribution map of scattering screens in the Galaxy.

\section*{Acknowledgements}

We thank Aditya Tamar for comments which have helped us to improve the manuscript.
In this work, we used the Astrogeo VLBI FITS image collection database, maintained by Leonid Petrov.
This database played a crucial role in our work, and we are profoundly grateful to all authors for their significant contributions in providing public VLBI data (see Section~\ref{s:data}).
The work of TAK was supported in the framework of the state assignment No. FFMR-2024-0010 of the Federal State Budget Scientific Institution ``Lebedev Physical Institute of the RAS''.
A.~V.~Plavin is a postdoctoral fellow at the Black Hole Initiative, which is funded by grants from the John Templeton Foundation (grants 60477, 61479, 62286) and the Gordon and Betty Moore Foundation (grant GBMF-8273).
The work of ABP was supported in the framework of the state assignment of the Federal State Budget Scientific Institution ``Crimean Astrophysical Observatory of RAS''.
YYK was supported by the MuSES project, which has received funding from the European Union (ERC grant agreement No 101142396). Views and opinions expressed are however those of the author(s) only and do not necessarily reflect those of the European Union or ERCEA. Neither the European Union nor the granting authority can be held responsible for them.
TAK thanks A.V.~Popkov for assistance with the development of the scripts used in this work.

\section*{Data availability}

The total intensity images of AGN jets used in our work are publicly available in the Astrogeo VLBI FITS image database collection
\href{https://doi.org/10.25966/kyy8-yp57}{DOI: 10.25966/kyy8-yp57}.
The dataset contains 19\,691 sources and approximately 100\,000 individual observations from 1994 to 2025. Details are provided in Section~\ref{s:data}.

\section*{Software}

For baseline-based fringe search, we used the FringeHunt.jl package\footnote{\url{https://github.com/aplavin/FringeHunt.jl}}.

\bibliographystyle{mnras}
\bibliography{article}

@ARTICLE{2025ApJS..276...38P,
       author = {{Petrov}, L.~Y. and {Kovalev}, Y.~Y.},
        title = "{The Radio Fundamental Catalog. I. Astrometry}",
      journal = {\apjs},
         year = 2025,
        month = feb,
       volume = {276},
       number = {2},
          eid = {38},
        pages = {38},
          doi = {10.3847/1538-4365/ad8c36},
archivePrefix = {arXiv},
       eprint = {2410.11794},
 primaryClass = {astro-ph.IM},
       adsurl = {https://ui.adsabs.harvard.edu/abs/2025ApJS..276...38P}
}

@ARTICLE{Kovalev05,
       author = {{Kovalev}, Y.~Y. and {Kellermann}, K.~I. and {Lister}, M.~L. and {Homan}, D.~C. and {Vermeulen}, R.~C. and {Cohen}, M.~H. and {Ros}, E. and {Kadler}, M. and {Lobanov}, A.~P. and {Zensus}, J.~A. and {Kardashev}, N.~S. and {Gurvits}, L.~I. and {Aller}, M.~F. and {Aller}, H.~D.},
        title = "{Sub-Milliarcsecond Imaging of Quasars and Active Galactic Nuclei. IV. Fine-Scale Structure}",
      journal = {\aj},
         year = 2005,
        month = dec,
       volume = {130},
       number = {6},
        pages = {2473-2505},
          doi = {10.1086/497430},
archivePrefix = {arXiv},
       eprint = {astro-ph/0505536},
 primaryClass = {astro-ph},
       adsurl = {https://ui.adsabs.harvard.edu/abs/2005AJ....130.2473K}
}

@ARTICLE{plavin2022,
       author = {{Plavin}, A.~V. and {Kovalev}, Y.~Y. and {Pushkarev}, A.~B.},
        title = "{Direction of Parsec-scale Jets for 9220 Active Galactic Nuclei}",
      journal = {\apjs},
         year = 2022,
        month = may,
       volume = {260},
       number = {1},
          eid = {4},
        pages = {4},
          doi = {10.3847/1538-4365/ac6352},
archivePrefix = {arXiv},
       eprint = {2203.13750},
 primaryClass = {astro-ph.HE},
       adsurl = {https://ui.adsabs.harvard.edu/abs/2022ApJS..260....4P}
}

@ARTICLE{SgrA_Johnson18,
      author = {{Johnson}, Michael D. and {Narayan}, Ramesh and {Psaltis}, Dimitrios and {Blackburn}, Lindy and {Kovalev}, Yuri Y. and {Gwinn}, Carl R. and {Zhao}, Guan
g-Yao and {Bower}, Geoffrey C. and {Moran}, James M. and {Kino}, Motoki and {Kramer}, Michael and {Akiyama}, Kazunori and {Dexter}, Jason and {Broderick}, Avery E. and {
Sironi}, Lorenzo},
       title = "{The Scattering and Intrinsic Structure of Sagittarius A* at Radio Wavelengths}",
     journal = {\apj},
        year = 2018,
       month = oct,
      volume = {865},
      number = {2},
         eid = {104},
       pages = {104},
         doi = {10.3847/1538-4357/aadcff},
archivePrefix = {arXiv},
      eprint = {1808.08966},
primaryClass = {astro-ph.GA},
      adsurl = {https://ui.adsabs.harvard.edu/abs/2018ApJ...865..104J}
}

@ARTICLE{Cordes86,
       author = {{Cordes}, J.~M. and {Pidwerbetsky}, A. and {Lovelace}, R.~V.~E.},
        title = "{Refractive and Diffractive Scattering in the Interstellar Medium}",
      journal = {\apj},
         year = 1986,
        month = nov,
       volume = {310},
        pages = {737},
          doi = {10.1086/164728},
       adsurl = {https://ui.adsabs.harvard.edu/abs/1986ApJ...310..737C}
}

@ARTICLE{Armstrong95,
       author = {{Armstrong}, J.~W. and {Rickett}, B.~J. and {Spangler}, S.~R.},
        title = "{Electron Density Power Spectrum in the Local Interstellar Medium}",
      journal = {\apj},
         year = 1995,
        month = apr,
       volume = {443},
        pages = {209},
          doi = {10.1086/175515},
       adsurl = {https://ui.adsabs.harvard.edu/abs/1995ApJ...443..209A}
}

@ARTICLE{BK79,
       author = {{Blandford}, R.~D. and {K{\"o}nigl}, A.},
        title = "{Relativistic jets as compact radio sources.}",
      journal = {\apj},
         year = 1979,
        month = aug,
       volume = {232},
        pages = {34-48},
          doi = {10.1086/157262},
       adsurl = {https://ui.adsabs.harvard.edu/abs/1979ApJ...232...34B}
}

@ARTICLE{Pushkarev15,
       author = {{Pushkarev}, A.~B. and {Kovalev}, Y.~Y.},
        title = "{Milky Way scattering properties and intrinsic sizes of active galactic nuclei cores probed by very long baseline interferometry surveys of compact extragalactic radio sources}",
      journal = {\mnras},
         year = 2015,
        month = oct,
       volume = {452},
       number = {4},
        pages = {4274-4282},
          doi = {10.1093/mnras/stv1539},
archivePrefix = {arXiv},
       eprint = {1507.02459},
 primaryClass = {astro-ph.HE},
       adsurl = {https://ui.adsabs.harvard.edu/abs/2015MNRAS.452.4274P}
}

@ARTICLE{Finkbeiner03,
       author = {{Finkbeiner}, Douglas P.},
        title = "{A Full-Sky H{\ensuremath{\alpha}} Template for Microwave Foreground Prediction}",
      journal = {\apjs},
         year = 2003,
        month = jun,
       volume = {146},
       number = {2},
        pages = {407-415},
          doi = {10.1086/374411},
archivePrefix = {arXiv},
       eprint = {astro-ph/0301558},
 primaryClass = {astro-ph},
       adsurl = {https://ui.adsabs.harvard.edu/abs/2003ApJS..146..407F}
}

@ARTICLE{2009JGeod..83..859P,
       author = {{Petrov}, Leonid and {Gordon}, David and {Gipson}, John and {MacMillan}, Dan and {Ma}, Chopo and {Fomalont}, Ed and {Walker}, R. Craig and {Carabajal}, Claudia},
        title = "{Precise geodesy with the Very Long Baseline Array}",
      journal = {Journal of Geodesy},
         year = 2009,
        month = sep,
       volume = {83},
       number = {9},
        pages = {859-876},
          doi = {10.1007/s00190-009-0304-7},
archivePrefix = {arXiv},
       eprint = {0806.0167},
 primaryClass = {physics.geo-ph},
       adsurl = {https://ui.adsabs.harvard.edu/abs/2009JGeod..83..859P}
}

@ARTICLE{2002ApJS..141...13B,
   author = {{Beasley}, A.~J. and {Gordon}, D. and {Peck}, A.~B. and {Petrov}, L. and
  {MacMillan}, D.~S. and {Fomalont}, E.~B. and {Ma}, C.},
    title = "{The VLBA Calibrator Survey-VCS1}",
  journal = {\apjs},
   eprint = {astro-ph/0201414},
     year = 2002,
    month = jul,
   volume = 141,
    pages = {13-21},
      doi = {10.1086/339806},
   adsurl = {http://adsabs.harvard.edu/abs/2002ApJS..141...13B}
}

@ARTICLE{2003AJ....126.2562F,
   author = {{Fomalont}, E.~B. and {Petrov}, L. and {MacMillan}, D.~S. and
  {Gordon}, D. and {Ma}, C.},
    title = "{The Second VLBA Calibrator Survey: VCS2}",
  journal = {\aj},
     year = 2003,
    month = nov,
   volume = 126,
    pages = {2562-2566},
      doi = {10.1086/378712},
   adsurl = {http://adsabs.harvard.edu/abs/2003AJ....126.2562F}
}

@ARTICLE{2005AJ....129.1163P,
   author = {{Petrov}, L. and {Kovalev}, Y.~Y. and {Fomalont}, E. and {Gordon}, D.
  },
    title = "{The Third VLBA Calibrator Survey: VCS3}",
  journal = {\aj},
   eprint = {astro-ph/0409698},
     year = 2005,
    month = feb,
   volume = 129,
    pages = {1163-1170},
      doi = {10.1086/426920},
   adsurl = {http://adsabs.harvard.edu/abs/2005AJ....129.1163P}
}

@ARTICLE{2006AJ....131.1872P,
       author = {{Petrov}, L. and {Kovalev}, Y.~Y. and {Fomalont}, E.~B. and {Gordon}, D.},
        title = "{The Fourth VLBA Calibrator Survey: VCS4}",
      journal = {\aj},
         year = 2006,
        month = mar,
       volume = {131},
       number = {3},
        pages = {1872-1879},
          doi = {10.1086/499947},
archivePrefix = {arXiv},
       eprint = {astro-ph/0508506},
 primaryClass = {astro-ph},
       adsurl = {https://ui.adsabs.harvard.edu/abs/2006AJ....131.1872P}
}

@ARTICLE{2007AJ....133.1236K,
   author = {{Kovalev}, Y.~Y. and {Petrov}, L. and {Fomalont}, E.~B. and
  {Gordon}, D.},
    title = "{The Fifth VLBA Calibrator Survey: VCS5}",
  journal = {\aj},
   eprint = {astro-ph/0607524},
     year = 2007,
    month = apr,
   volume = 133,
    pages = {1236-1242},
      doi = {10.1086/511157},
   adsurl = {http://adsabs.harvard.edu/abs/2007AJ....133.1236K}
}

@ARTICLE{2008AJ....136..580P,
       author = {{Petrov}, L. and {Kovalev}, Y.~Y. and {Fomalont}, E.~B. and {Gordon}, D.},
        title = "{The Sixth VLBA Calibrator Survey: VCS6}",
      journal = {\aj},
         year = 2008,
        month = aug,
       volume = {136},
       number = {2},
        pages = {580-585},
          doi = {10.1088/0004-6256/136/2/580},
archivePrefix = {arXiv},
       eprint = {0801.3895},
 primaryClass = {astro-ph},
       adsurl = {https://ui.adsabs.harvard.edu/abs/2008AJ....136..580P}
}

@ARTICLE{2011AJ....142...35P,
       author = {{Petrov}, L. and {Kovalev}, Y.~Y. and {Fomalont}, E.~B. and {Gordon}, D.},
        title = "{The Very Long Baseline Array Galactic Plane Survey{\textemdash}VGaPS}",
      journal = {\aj},
         year = "2011",
        month = "Aug",
       volume = {142},
       number = {2},
          eid = {35},
        pages = {35},
          doi = {10.1088/0004-6256/142/2/35},
archivePrefix = {arXiv},
       eprint = {1101.1460},
 primaryClass = {astro-ph.CO},
       adsurl = {https://ui.adsabs.harvard.edu/abs/2011AJ....142...35P}
}

@ARTICLE{2012MNRAS.419.1097P,
       author = {{Petrov}, Leonid},
        title = "{The EVN Galactic Plane Survey - EGaPS}",
      journal = {\mnras},
         year = "2012",
        month = "Jan",
       volume = {419},
       number = {2},
        pages = {1097-1106},
          doi = {10.1111/j.1365-2966.2011.19765.x},
archivePrefix = {arXiv},
       eprint = {1106.4883},
 primaryClass = {astro-ph.CO},
       adsurl = {https://ui.adsabs.harvard.edu/abs/2012MNRAS.419.1097P}
}

@ARTICLE{2011AJ....142..105P,
       author = {{Petrov}, L.},
        title = "{The Catalog of Positions of Optically Bright Extragalactic Radio Sources OBRS-1}",
      journal = {\aj},
         year = "2011",
        month = "Oct",
       volume = {142},
       number = {4},
          eid = {105},
        pages = {105},
          doi = {10.1088/0004-6256/142/4/105},
archivePrefix = {arXiv},
       eprint = {1103.2840},
 primaryClass = {astro-ph.CO},
       adsurl = {https://ui.adsabs.harvard.edu/abs/2011AJ....142..105P}
}

@ARTICLE{2013AJ....146....5P,
       author = {{Petrov}, L.},
        title = "{The Catalog of Positions of Optically Bright Extragalactic Radio Sources OBRS-2}",
      journal = {\aj},
         year = "2013",
        month = "Jul",
       volume = {146},
       number = {1},
          eid = {5},
        pages = {5},
          doi = {10.1088/0004-6256/146/1/5},
archivePrefix = {arXiv},
       eprint = {1301.5407},
 primaryClass = {astro-ph.IM},
       adsurl = {https://ui.adsabs.harvard.edu/abs/2013AJ....146....5P}
}

@ARTICLE{2011MNRAS.414.2528P,
       author = {{Petrov}, Leonid and {Phillips}, Chris and {Bertarini}, Alessandra and
         {Murphy}, Tara and {Sadler}, Elaine M.},
        title = "{The LBA Calibrator Survey of southern compact extragalactic radio sources - LCS1}",
      journal = {\mnras},
         year = "2011",
        month = "Jul",
       volume = {414},
       number = {3},
        pages = {2528-2539},
          doi = {10.1111/j.1365-2966.2011.18570.x},
archivePrefix = {arXiv},
       eprint = {1012.2607},
 primaryClass = {astro-ph.CO},
       adsurl = {https://ui.adsabs.harvard.edu/abs/2011MNRAS.414.2528P}
}

@ARTICLE{2019MNRAS.485...88P,
       author = {{Petrov}, Leonid and {de Witt}, Alet and {Sadler}, Elaine M. and
         {Phillips}, Chris and {Horiuchi}, Shinji},
        title = "{The Second LBA Calibrator Survey of southern compact extragalactic radio sources - LCS2}",
      journal = {\mnras},
         year = "2019",
        month = "May",
       volume = {485},
       number = {1},
        pages = {88-101},
          doi = {10.1093/mnras/stz242},
archivePrefix = {arXiv},
       eprint = {1812.02916},
 primaryClass = {astro-ph.IM},
       adsurl = {https://ui.adsabs.harvard.edu/abs/2019MNRAS.485...88P}
}

@ARTICLE{2015ApJS..217....4S,
       author = {{Schinzel}, Frank K. and {Petrov}, Leonid and {Taylor}, Gregory B. and
         {Mahony}, Elizabeth K. and {Edwards}, Philip G. and {Kovalev}, Yuri Y.},
        title = "{New Associations of Gamma-Ray Sources from the Fermi Second Source Catalog}",
      journal = {\apjs},
         year = "2015",
        month = "Mar",
       volume = {217},
       number = {1},
          eid = {4},
        pages = {4},
          doi = {10.1088/0067-0049/217/1/4},
archivePrefix = {arXiv},
       eprint = {1408.6217},
 primaryClass = {astro-ph.HE},
       adsurl = {https://ui.adsabs.harvard.edu/abs/2015ApJS..217....4S}
}

@ARTICLE{2017ApJS..230...13S,
       author = {{Shu}, Fengchun and {Petrov}, Leonid and {Jiang}, Wu and {Xia}, Bo and
         {Jiang}, Tianyu and {Cui}, Yuzhu and {Takefuji}, Kazuhiro and
         {McCallum}, Jamie and {Lovell}, Jim and {Yi}, Sang-oh and
         {Hao}, Longfei and {Yang}, Wenjun and {Zhang}, Hua and {Chen}, Zhong and
         {Li}, Jinling},
        title = "{VLBI Ecliptic Plane Survey: VEPS-1}",
      journal = {\apjs},
         year = "2017",
        month = "Jun",
       volume = {230},
       number = {2},
          eid = {13},
        pages = {13},
          doi = {10.3847/1538-4365/aa71a3},
archivePrefix = {arXiv},
       eprint = {1701.07287},
 primaryClass = {astro-ph.IM},
       adsurl = {https://ui.adsabs.harvard.edu/abs/2017ApJS..230...13S}
}

@ARTICLE{2021AJ....161...14P,
       author = {{Petrov}, Leonid},
        title = "{The Wide-field VLBA Calibrator Survey: WFCS}",
      journal = {\aj},
         year = 2021,
        month = jan,
       volume = {161},
       number = {1},
          eid = {14},
        pages = {14},
          doi = {10.3847/1538-3881/abc4e1},
archivePrefix = {arXiv},
       eprint = {2008.09243},
 primaryClass = {astro-ph.IM},
       adsurl = {https://ui.adsabs.harvard.edu/abs/2021AJ....161...14P}
}

@ARTICLE{2021AJ....161...88P,
       author = {{Popkov}, A.~V. and {Kovalev}, Y.~Y. and {Petrov}, L.~Y. and {Kovalev}, Yu. A.},
        title = "{Parsec-scale Properties of Steep- and Flat-spectrum Extragalactic Radio Sources from a VLBA Survey of a Complete North Polar Cap Sample}",
      journal = {\aj},
         year = 2021,
        month = feb,
       volume = {161},
       number = {2},
          eid = {88},
        pages = {88},
          doi = {10.3847/1538-3881/abd18c},
archivePrefix = {arXiv},
       eprint = {2008.06803},
 primaryClass = {astro-ph.GA},
       adsurl = {https://ui.adsabs.harvard.edu/abs/2021AJ....161...88P}
}

@ARTICLE{2012A&A...544A..34P,
   author = {{Pushkarev}, A.~B. and {Kovalev}, Y.~Y.},
    title = "{Single-epoch VLBI imaging study of bright active galactic nuclei at 2 GHz and 8 GHz}",
  journal = {\aap},
archivePrefix = "arXiv",
   eprint = {1205.5559},
     year = 2012,
    month = aug,
   volume = 544,
      eid = {A34},
    pages = {A34},
      doi = {10.1051/0004-6361/201219352},
   adsurl = {http://adsabs.harvard.edu/abs/2012A%26A...544A..34P}
}

@ARTICLE{2012ApJ...758...84P,
   author = {{Piner}, B.~G. and {Pushkarev}, A.~B. and {Kovalev}, Y.~Y. and
  {Marvin}, C.~J. and {Arenson}, J.~G. and {Charlot}, P. and {Fey}, A.~L. and
  {Collioud}, A. and {Voitsik}, P.~A.},
    title = "{Relativistic Jets in the Radio Reference Frame Image Database. II. Blazar Jet Accelerations from the First 10 Years of Data (1994-2003)}",
  journal = {\apj},
archivePrefix = "arXiv",
   eprint = {1208.4399},
     year = 2012,
    month = oct,
   volume = 758,
      eid = {84},
    pages = {84},
      doi = {10.1088/0004-637X/758/2/84},
   adsurl = {http://adsabs.harvard.edu/abs/2012ApJ...758...84P}
}

@ARTICLE{2007ApJ...658..203H,
       author = {{Helmboldt}, J.~F. and {Taylor}, G.~B. and {Tremblay}, S. and {Fassnacht}, C.~D. and {Walker}, R.~C. and {Myers}, S.~T. and {Sjouwerman}, L.~O. and {Pearson}, T.~J. and {Readhead}, A.~C.~S. and {Weintraub}, L. and {Gehrels}, N. and {Romani}, R.~W. and {Healey}, S. and {Michelson}, P.~F. and {Blandford}, R.~D. and {Cotter}, G.},
        title = "{The VLBA Imaging and Polarimetry Survey at 5 GHz}",
      journal = {\apj},
         year = 2007,
        month = mar,
       volume = {658},
       number = {1},
        pages = {203-216},
          doi = {10.1086/511005},
archivePrefix = {arXiv},
       eprint = {astro-ph/0611459},
 primaryClass = {astro-ph},
       adsurl = {https://ui.adsabs.harvard.edu/abs/2007ApJ...658..203H}
}

@ARTICLE{2008AJ....136..159L,
       author = {{Lee}, Sang-Sung and {Lobanov}, Andrei P. and {Krichbaum}, Thomas P. and {Witzel}, Arno and {Zensus}, Anton and {Bremer}, Michael and {Greve}, Albert and {Grewing}, Michael},
        title = "{A Global 86 GHz VLBI Survey of Compact Radio Sources}",
      journal = {\aj},
         year = 2008,
        month = jul,
       volume = {136},
       number = {1},
        pages = {159-180},
          doi = {10.1088/0004-6256/136/1/159},
archivePrefix = {arXiv},
       eprint = {0803.4035},
 primaryClass = {astro-ph},
       adsurl = {https://ui.adsabs.harvard.edu/abs/2008AJ....136..159L}
}

@ARTICLE{2019A&A...622A..92N,
       author = {{Nair}, Dhanya G. and {Lobanov}, Andrei P. and {Krichbaum}, Thomas P. and {Ros}, Eduardo and {Zensus}, Johann Anton and {Kovalev}, Yuri Y. and {Lee}, Sang-Sung and {Mertens}, Florent and {Hagiwara}, Yoshiaki and {Bremer}, Michael and {Lindqvist}, Michael and {de Vicente}, Pablo},
        title = "{Global millimeter VLBI array survey of ultracompact extragalactic radio sources at 86 GHz}",
      journal = {\aap},
         year = 2019,
        month = feb,
       volume = {622},
          eid = {A92},
        pages = {A92},
          doi = {10.1051/0004-6361/201833122},
archivePrefix = {arXiv},
       eprint = {1808.09243},
 primaryClass = {astro-ph.GA},
       adsurl = {https://ui.adsabs.harvard.edu/abs/2019A&A...622A..92N}
}

@ARTICLE{2017ApJ...846...98J,
       author = {{Jorstad}, Svetlana G. and {Marscher}, Alan P. and {Morozova}, Daria A. and {Troitsky}, Ivan S. and {Agudo}, Iv{\'a}n and {Casadio}, Carolina and {Foord}, Adi and {G{\'o}mez}, Jos{\'e} L. and {MacDonald}, Nicholas R. and {Molina}, Sol N. and {L{\"a}hteenm{\"a}ki}, Anne and {Tammi}, Joni and {Tornikoski}, Merja},
        title = "{Kinematics of Parsec-scale Jets of Gamma-Ray Blazars at 43 GHz within the VLBA-BU-BLAZAR Program}",
      journal = {\apj},
         year = 2017,
        month = sep,
       volume = {846},
       number = {2},
          eid = {98},
        pages = {98},
          doi = {10.3847/1538-4357/aa8407},
archivePrefix = {arXiv},
       eprint = {1711.03983},
 primaryClass = {astro-ph.GA},
       adsurl = {https://ui.adsabs.harvard.edu/abs/2017ApJ...846...98J}
}

@ARTICLE{Elmegreen_2004,
       author = {{Elmegreen}, Bruce G. and {Scalo}, John},
        title = "{Interstellar Turbulence I: Observations and Processes}",
      journal = {\araa},
         year = 2004,
        month = sep,
       volume = {42},
       number = {1},
        pages = {211-273},
          doi = {10.1146/annurev.astro.41.011802.094859},
archivePrefix = {arXiv},
       eprint = {astro-ph/0404451},
 primaryClass = {astro-ph},
       adsurl = {https://ui.adsabs.harvard.edu/abs/2004ARA&A..42..211E}
}

@ARTICLE{Scalo2004,
       author = {{Scalo}, John and {Elmegreen}, Bruce G.},
        title = "{Interstellar Turbulence II: Implications and Effects}",
      journal = {\araa},
         year = 2004,
        month = sep,
       volume = {42},
       number = {1},
        pages = {275-316},
          doi = {10.1146/annurev.astro.42.120403.143327},
archivePrefix = {arXiv},
       eprint = {astro-ph/0404452},
 primaryClass = {astro-ph},
       adsurl = {https://ui.adsabs.harvard.edu/abs/2004ARA&A..42..275S}
}

@ARTICLE{Rickett1990,
       author = {{Rickett}, B.~J.},
        title = "{Radio propagation through the turbulent interstellar plasma.}",
      journal = {\araa},
         year = 1990,
        month = jan,
       volume = {28},
        pages = {561-605},
          doi = {10.1146/annurev.aa.28.090190.003021},
       adsurl = {https://ui.adsabs.harvard.edu/abs/1990ARA&A..28..561R}
}

@article{Kolmogorov41,
   author = {{Kolmogorov}, A.N.},
   title="{The local structure of turbulence in incompressible viscous fluid for very large Reynolds numbers.}",
   pages={299-303},
   journal={Proceedings of the USSR Academy of Sciences},
   year={1941},
}

@ARTICLE{Spangler1990_1,
       author = {{Spangler}, Steven R. and {Gwinn}, Carl R.},
        title = "{Evidence for an Inner Scale to the Density Turbulence in the Interstellar Medium}",
      journal = {\apjl},
         year = 1990,
        month = apr,
       volume = {353},
        pages = {L29},
          doi = {10.1086/185700},
       adsurl = {https://ui.adsabs.harvard.edu/abs/1990ApJ...353L..29S}
}

@ARTICLE{Konigl1981,
       author = {{K{\"o}nigl}, A.},
        title = "{Relativistic jets as X-ray and gamma-ray sources.}",
      journal = {\apj},
         year = 1981,
        month = feb,
       volume = {243},
        pages = {700-709},
          doi = {10.1086/158638},
       adsurl = {https://ui.adsabs.harvard.edu/abs/1981ApJ...243..700K}
}

@ARTICLE{Pushkarev2013,
       author = {{Pushkarev}, A.~B. and {Kovalev}, Y.~Y. and {Lister}, M.~L. and {Hovatta}, T. and {Savolainen}, T. and {Aller}, M.~F. and {Aller}, H.~D. and {Ros}, E. and {Zensus}, J.~A. and {Richards}, J.~L. and {Max-Moerbeck}, W. and {Readhead}, A.~C.~S.},
        title = "{VLBA observations of a rare multiple quasar imaging event caused by refraction in the interstellar medium}",
      journal = {\aap},
         year = 2013,
        month = jul,
       volume = {555},
          eid = {A80},
        pages = {A80},
          doi = {10.1051/0004-6361/201321484},
archivePrefix = {arXiv},
       eprint = {1305.6005},
 primaryClass = {astro-ph.CO},
       adsurl = {https://ui.adsabs.harvard.edu/abs/2013A&A...555A..80P}
}

@ARTICLE{Johnson2016,
       author = {{Johnson}, Michael D. and {Kovalev}, Yuri Y. and {Gwinn}, Carl R. and {Gurvits}, Leonid I. and {Narayan}, Ramesh and {Macquart}, Jean-Pierre and {Jauncey}, David L. and {Voitsik}, Peter A. and {Anderson}, James M. and {Sokolovsky}, Kirill V. and {Lisakov}, Mikhail M.},
        title = "{Extreme Brightness Temperatures and Refractive Substructure in 3C273 with RadioAstron}",
      journal = {\apjl},
         year = 2016,
        month = mar,
       volume = {820},
       number = {1},
          eid = {L10},
        pages = {L10},
          doi = {10.3847/2041-8205/820/1/L10},
archivePrefix = {arXiv},
       eprint = {1601.05810},
 primaryClass = {astro-ph.HE},
       adsurl = {https://ui.adsabs.harvard.edu/abs/2016ApJ...820L..10J}
}

@ARTICLE{Duffett_Smith1976,
       author = {{Duffett-Smith}, P.~J. and {Readhead}, A.~C.~S.},
        title = "{The angular broadening of radio sources by scattering in the interstellar medium.}",
      journal = {\mnras},
         year = 1976,
        month = jan,
       volume = {174},
        pages = {7-17},
          doi = {10.1093/mnras/174.1.7},
       adsurl = {https://ui.adsabs.harvard.edu/abs/1976MNRAS.174....7D}
}

@ARTICLE{Koryukova2022,
       author = {{Koryukova}, T.~A. and {Pushkarev}, A.~B. and {Plavin}, A.~V. and {Kovalev}, Y.~Y.},
        title = "{Tracing Milky Way scattering by compact extragalactic radio sources}",
      journal = {\mnras},
         year = 2022,
        month = sep,
       volume = {515},
       number = {2},
        pages = {1736-1750},
          doi = {10.1093/mnras/stac1898},
archivePrefix = {arXiv},
       eprint = {2201.04359},
 primaryClass = {astro-ph.GA},
       adsurl = {https://ui.adsabs.harvard.edu/abs/2022MNRAS.515.1736K}
}

@INPROCEEDINGS{Greisen2003,
       author = {{Greisen}, E.~W.},
        title = "{AIPS, the VLA, and the VLBA}",
    booktitle = {Information Handling in Astronomy - Historical Vistas},
         year = 2003,
       editor = {{Heck}, Andr{\'e}},
       series = {Astrophysics and Space Science Library},
       volume = {285},
        month = mar,
        pages = {109},
          doi = {10.1007/0-306-48080-8_7},
       adsurl = {https://ui.adsabs.harvard.edu/abs/2003ASSL..285..109G}
}

@ARTICLE{Gabani2006,
       author = {{Gab{\'a}nyi}, K. {\'E}. and {Krichbaum}, T.~P. and {Britzen}, S. and {Bach}, U. and {Ros}, E. and {Witzel}, A. and {Zensus}, J.~A.},
        title = "{High frequency VLBI observations of the scatter-broadened quasar B 2005+403}",
      journal = {\aap},
         year = 2006,
        month = may,
       volume = {451},
       number = {1},
        pages = {85-98},
          doi = {10.1051/0004-6361:20054017},
archivePrefix = {arXiv},
       eprint = {astro-ph/0601362},
 primaryClass = {astro-ph},
       adsurl = {https://ui.adsabs.harvard.edu/abs/2006A&A...451...85G}
}

@ARTICLE{Fey1989,
       author = {{Fey}, Alan L. and {Spangler}, Steven R. and {Mutel}, Robert L.},
        title = "{VLBI Angular Broadening Measurements in the Cygnus Region}",
      journal = {\apj},
         year = 1989,
        month = feb,
       volume = {337},
        pages = {730},
          doi = {10.1086/167144},
       adsurl = {https://ui.adsabs.harvard.edu/abs/1989ApJ...337..730F}
}

@ARTICLE{Koryukova2023,
       author = {{Koryukova}, T.~A. and {Pushkarev}, A.~B. and {Kiehlmann}, S. and {Readhead}, A.~C.~S.},
        title = "{Multiple imaging of the quasar 2005 + 403 formed by anisotropic scattering}",
      journal = {\mnras},
         year = 2023,
        month = dec,
       volume = {526},
       number = {4},
        pages = {5932-5948},
          doi = {10.1093/mnras/stad3052},
archivePrefix = {arXiv},
       eprint = {2308.15274},
 primaryClass = {astro-ph.GA},
       adsurl = {https://ui.adsabs.harvard.edu/abs/2023MNRAS.526.5932K}
}

@ARTICLE{Lister2021,
       author = {{Lister}, M.~L. and {Homan}, D.~C. and {Kellermann}, K.~I. and {Kovalev}, Y.~Y. and {Pushkarev}, A.~B. and {Ros}, E. and {Savolainen}, T.},
        title = "{Monitoring Of Jets in Active Galactic Nuclei with VLBA Experiments. XVIII. Kinematics and Inner Jet Evolution of Bright Radio-loud Active Galaxies}",
      journal = {\apj},
         year = 2021,
        month = dec,
       volume = {923},
       number = {1},
          eid = {30},
        pages = {30},
          doi = {10.3847/1538-4357/ac230f},
archivePrefix = {arXiv},
       eprint = {2108.13358},
 primaryClass = {astro-ph.HE},
       adsurl = {https://ui.adsabs.harvard.edu/abs/2021ApJ...923...30L}
}

@ARTICLE{Plavin2026,
       author = {{Plavin}, A.~V. and {Pushkarev}, A.~B. and {Kovalev}, Y.~Y.},
        title = "{Direct Very Long Baseline Interferometry Detection of Interstellar Turbulence Imprint on a Quasar: TXS 2005+403}",
      journal = {\apjl},
         year = 2026,
        month = may,
       volume = {1003},
       number = {1},
          eid = {L4},
        pages = {L4},
          doi = {10.3847/2041-8213/ae60f4},
archivePrefix = {arXiv},
       eprint = {2602.24255},
 primaryClass = {astro-ph.GA},
       adsurl = {https://ui.adsabs.harvard.edu/abs/2026ApJ..1003L...4P}
}

@ARTICLE{JohnsonGwinn2015,
       author = {{Johnson}, Michael D. and {Gwinn}, Carl R.},
        title = "{Theory and Simulations of Refractive Substructure in Resolved Scatter-broadened Images}",
      journal = {\apj},
         year = 2015,
        month = jun,
       volume = {805},
       number = {2},
          eid = {180},
        pages = {180},
          doi = {10.1088/0004-637X/805/2/180},
archivePrefix = {arXiv},
       eprint = {1502.05722},
 primaryClass = {astro-ph.IM},
       adsurl = {https://ui.adsabs.harvard.edu/abs/2015ApJ...805..180J}
}

@ARTICLE{Pilipenko2018,
       author = {{Pilipenko}, S.~V. and {Kovalev}, Y.~Y. and {Andrianov}, A.~S. and {Bach}, U. and {Buttaccio}, S. and {Cassaro}, P. and {Cim{\`o}}, G. and {Edwards}, P.~G. and {Gawro{\'n}ski}, M.~P. and {Gurvits}, L.~I. and {Hovatta}, T. and {Jauncey}, D.~L. and {Johnson}, M.~D. and {Kovalev}, Yu A. and {Kutkin}, A.~M. and {Lisakov}, M.~M. and {Melnikov}, A.~E. and {Orlati}, A. and {Rudnitskiy}, A.~G. and {Sokolovsky}, K.~V. and {Stanghellini}, C. and {de Vicente}, P. and {Voitsik}, P.~A. and {Wolak}, P. and {Zhekanis}, G.~V.},
        title = "{The high brightness temperature of B0529+483 revealed by RadioAstron and implications for interstellar scattering}",
      journal = {\mnras},
         year = 2018,
        month = mar,
       volume = {474},
       number = {3},
        pages = {3523-3534},
          doi = {10.1093/mnras/stx2991},
archivePrefix = {arXiv},
       eprint = {1711.06713},
 primaryClass = {astro-ph.HE},
       adsurl = {https://ui.adsabs.harvard.edu/abs/2018MNRAS.474.3523P}
}

@ARTICLE{Gwinn2014,
       author = {{Gwinn}, C.~R. and {Kovalev}, Y.~Y. and {Johnson}, M.~D. and {Soglasnov}, V.~A.},
        title = "{Discovery of Substructure in the Scatter-broadened Image of Sgr A*}",
      journal = {\apjl},
         year = 2014,
        month = oct,
       volume = {794},
       number = {1},
          eid = {L14},
        pages = {L14},
          doi = {10.1088/2041-8205/794/1/L14},
archivePrefix = {arXiv},
       eprint = {1409.0530},
 primaryClass = {astro-ph.GA},
       adsurl = {https://ui.adsabs.harvard.edu/abs/2014ApJ...794L..14G}
}

@book{Efron1993,
  title={An Introduction to the Bootstrap},
  author={Efron, Bradley and Tibshirani, Robert J.},
  year={1993},
  publisher={Chapman \& Hall/CRC}
}

@ARTICLE{Narayan1989,
       author = {{Narayan}, Ramesh and {Goodman}, Jeremy},
        title = "{The shape of a scatter-broadened image - I. Numerical simulations and physical principles.}",
      journal = {\mnras},
         year = 1989,
        month = jun,
       volume = {238},
        pages = {963-1028},
          doi = {10.1093/mnras/238.3.963},
       adsurl = {https://ui.adsabs.harvard.edu/abs/1989MNRAS.238..963N}
}

@ARTICLE{Johnson2021,
       author = {{Johnson}, Michael D. and {Kovalev}, Yuri Y. and {Lisakov}, Mikhail M. and {Voitsik}, Petr A. and {Gwinn}, Carl R. and {Bruni}, Gabriele},
        title = "{First Space-VLBI Observations of Sagittarius A*}",
      journal = {\apjl},
         year = 2021,
        month = dec,
       volume = {922},
       number = {2},
          eid = {L28},
        pages = {L28},
          doi = {10.3847/2041-8213/ac3917},
archivePrefix = {arXiv},
       eprint = {2111.06423},
 primaryClass = {astro-ph.GA},
       adsurl = {https://ui.adsabs.harvard.edu/abs/2021ApJ...922L..28J}
}

@ARTICLE{Pushkarev2026,
       author = {{Pushkarev}, A.~B. and {Brunthaler}, A. and {Kovalev}, Y.~Y. and {Lisakov}, M.~M. and {Pashchenko}, I.~N. and {Plavin}, A.~V. and {Roy}, N. and {Voitsik}, P.~A. and {Dzib}, S.~A. and {Koryukova}, T.~A. and {Yang}, A.~Y.},
        title = "{Heavy interstellar scattering toward the near end of the Galactic bar}",
      journal = {\mnras},
         year = 2026,
        month = may,
       volume = {548},
       number = {2},
          eid = {stag655},
        pages = {stag655},
          doi = {10.1093/mnras/stag655},
archivePrefix = {arXiv},
       eprint = {2602.06785},
 primaryClass = {astro-ph.GA},
       adsurl = {https://ui.adsabs.harvard.edu/abs/2026MNRAS.548ag655P}
}

@ARTICLE{Rickett2009,
       author = {{Rickett}, Barney and {Johnston}, Simon and {Tomlinson}, Taryn and {Reynolds}, John},
        title = "{The inner scale of the plasma turbulence towards PSR J1644-4559}",
      journal = {\mnras},
         year = 2009,
        month = may,
       volume = {395},
       number = {3},
        pages = {1391-1402},
          doi = {10.1111/j.1365-2966.2009.14471.x},
       adsurl = {https://ui.adsabs.harvard.edu/abs/2009MNRAS.395.1391R}
}

@INPROCEEDINGS{Selina2018,
       author = {{Selina}, Robert J. and {Murphy}, Eric J. and {McKinnon}, Mark and {Beasley}, Anthony and {Butler}, Bryan and {Carilli}, Chris and {Clark}, Barry and {Erickson}, Alan and {Grammer}, Wes and {Jackson}, James and {Kent}, Brian and {Mason}, Brian and {Morgan}, Matthew and {Ojeda}, Omar and {Shillue}, William and {Sturgis}, Silver and {Urbain}, Denis},
        title = "{The Next-Generation Very Large Array: a technical overview}",
    booktitle = {Ground-based and Airborne Telescopes VII},
         year = 2018,
       editor = {{Marshall}, Heather K. and {Spyromilio}, Jason},
       series = {Society of Photo-Optical Instrumentation Engineers (SPIE) Conference Series},
       volume = {10700},
        month = jul,
          eid = {107001O},
        pages = {107001O},
          doi = {10.1117/12.2312089},
archivePrefix = {arXiv},
       eprint = {1806.08405},
 primaryClass = {astro-ph.IM},
       adsurl = {https://ui.adsabs.harvard.edu/abs/2018SPIE10700E..1OS}
}

@ARTICLE{Kovalev2026,
       author = {{Kovalev}, Y.~Y. and {Bruni}, G. and {An}, T. and {Bignall}, H.~E. and {Edwards}, P.~G. and {Garcia-Miro}, C. and {Giroletti}, M. and {Gurvits}, L.~I. and {Kadler}, M. and {Kim}, J.-Y. and {Kravchenko}, E.~V. and {Liodakis}, I. and {Liu}, Y.~Q. and {Plavin}, A.~V. and {Popkov}, A.~V. and {Pushkarev}, A.~B. and {Savolainen}, T. and {Shen}, Z.-Q. and {Tamar}, A. and {Traianou}, E.},
        title = "{Unique Science Opportunities for Space VLBI Systems with the SKA Telescopes}",
      journal = {arXiv e-prints},
         year = 2026,
        month = jun,
          eid = {arXiv:2606.25137},
        pages = {arXiv:2606.25137},
          doi = {10.48550/arXiv.2606.25137},
archivePrefix = {arXiv},
       eprint = {2606.25137},
 primaryClass = {astro-ph.HE},
       adsurl = {https://ui.adsabs.harvard.edu/abs/2026arXiv260625137K}
}

@ARTICLE{Goodman1989,
       author = {{Goodman}, J. and {Narayan}, R.},
        title = "{The Shape of a Scatter Broadened Image - II. Interferometric Visibilities}",
      journal = {\mnras},
         year = 1989,
        month = jun,
       volume = {238},
        pages = {995},
          doi = {10.1093/mnras/238.3.995},
       adsurl = {https://ui.adsabs.harvard.edu/abs/1989MNRAS.238..995G}
}

@ARTICLE{Kovalev2016,
       author = {{Kovalev}, Y.~Y. and {Kardashev}, N.~S. and {Kellermann}, K.~I. and {Lobanov}, A.~P. and {Johnson}, M.~D. and {Gurvits}, L.~I. and {Voitsik}, P.~A. and {Zensus}, J.~A. and {Anderson}, J.~M. and {Bach}, U. and {Jauncey}, D.~L. and {Ghigo}, F. and {Ghosh}, T. and {Kraus}, A. and {Kovalev}, Yu. A. and {Lisakov}, M.~M. and {Petrov}, L. Yu. and {Romney}, J.~D. and {Salter}, C.~J. and {Sokolovsky}, K.~V.},
        title = "{RadioAstron Observations of the Quasar 3C273: A Challenge to the Brightness Temperature Limit}",
      journal = {\apjl},
         year = 2016,
        month = mar,
       volume = {820},
       number = {1},
          eid = {L9},
        pages = {L9},
          doi = {10.3847/2041-8205/820/1/L9},
archivePrefix = {arXiv},
       eprint = {1601.05806},
 primaryClass = {astro-ph.HE},
       adsurl = {https://ui.adsabs.harvard.edu/abs/2016ApJ...820L...9K}
}

@ARTICLE{Koryukova2025,
       author = {{Koryukova}, T.~A. and {Trushkin}, S.~A. and {Pashchenko}, I.~N. and {Pushkarev}, A.~B.},
        title = "{Probing plasma scattering screens towards the quasar 2005 + 403 with long-term RATAN-600 observations}",
      journal = {\mnras},
         year = 2025,
        month = oct,
       volume = {542},
       number = {4},
        pages = {2733-2751},
          doi = {10.1093/mnras/staf1369},
archivePrefix = {arXiv},
       eprint = {2503.22858},
 primaryClass = {astro-ph.GA},
       adsurl = {https://ui.adsabs.harvard.edu/abs/2025MNRAS.542.2733K}
}

@ARTICLE{Issaoun2019,
       author = {{Issaoun}, S. and {Johnson}, M.~D. and {Blackburn}, L. and {Brinkerink}, C.~D. and {Mo{\'s}cibrodzka}, M. and {Chael}, A. and {Goddi}, C. and {Mart{\'\i}-Vidal}, I. and {Wagner}, J. and {Doeleman}, S.~S. and {Falcke}, H. and {Krichbaum}, T.~P. and {Akiyama}, K. and {Bach}, U. and {Bouman}, K.~L. and {Bower}, G.~C. and {Broderick}, A. and {Cho}, I. and {Crew}, G. and {Dexter}, J. and {Fish}, V. and {Gold}, R. and {G{\'o}mez}, J.~L. and {Hada}, K. and {Hern{\'a}ndez-G{\'o}mez}, A. and {Jan{\ss}en}, M. and {Kino}, M. and {Kramer}, M. and {Loinard}, L. and {Lu}, R.-S. and {Markoff}, S. and {Marrone}, D.~P. and {Matthews}, L.~D. and {Moran}, J.~M. and {M{\"u}ller}, C. and {Roelofs}, F. and {Ros}, E. and {Rottmann}, H. and {Sanchez}, S. and {Tilanus}, R.~P.~J. and {de Vicente}, P. and {Wielgus}, M. and {Zensus}, J.~A. and {Zhao}, G.-Y.},
        title = "{The Size, Shape, and Scattering of Sagittarius A* at 86 GHz: First VLBI with ALMA}",
      journal = {\apj},
         year = 2019,
        month = jan,
       volume = {871},
       number = {1},
          eid = {30},
        pages = {30},
          doi = {10.3847/1538-4357/aaf732},
archivePrefix = {arXiv},
       eprint = {1901.06226},
 primaryClass = {astro-ph.HE},
       adsurl = {https://ui.adsabs.harvard.edu/abs/2019ApJ...871...30I}
}

@ARTICLE{Lobanov2015,
       author = {{Lobanov}, Andrei},
        title = "{Brightness temperature constraints from interferometric visibilities}",
      journal = {\aap},
         year = 2015,
        month = feb,
       volume = {574},
          eid = {A84},
        pages = {A84},
          doi = {10.1051/0004-6361/201425084},
archivePrefix = {arXiv},
       eprint = {1412.2121},
 primaryClass = {astro-ph.IM},
       adsurl = {https://ui.adsabs.harvard.edu/abs/2015A&A...574A..84L}
}

\appendix

\section{Angular size estimations for eight scattered AGNs}
\label{s:appendix_sizes}

In this section we present our measurements for the eight scattered AGNs identified as sources with significant refractive substructure signatures. Table~\ref{tab:appendix_epochs_part1} lists the flux density and angular size obtained from single Gaussian fits of the visibility data for each source and observing epoch. The table lists the estimated model parameters and only their statistical errors.

\begin{table}
\centering
\caption{Information on sources with significant scattering substructure: observing epochs, frequencies, flux densities and angular sizes defined as the full width at half maximum of fitted Gaussian component.}
\small
\setlength{\tabcolsep}{4pt}
\begin{tabular}{lcccc}
\hline
J2000 name & Epoch & $\nu$ (GHz) & $S$ (mJy) & $\theta$ (mas) \\
\hline
J0532$+$0732 & 1995-07-15 & 2.3 & $3261.8 \pm 51.3$ & $33.10 \pm 0.29$ \\
 & 1997-01-11 & 2.3 & $2040.7 \pm 14.2$ & $30.46 \pm 0.12$ \\
 & 2018-12-04 & 2.3 & $967.6 \pm 6.8$ & $26.44 \pm 0.12$ \\
 & 2021-08-19 & 4.2 & $853.5 \pm 3.4$ & $5.80 \pm 0.02$ \\ 
 & 2018-10-24 & 4.4 & $860.6 \pm 3.1$ & $7.02 \pm 0.03$ \\
 & 1996-06-05 & 4.9 & $1795.5 \pm 8.4$ & $5.82 \pm 0.02$ \\
 & 2018-10-24 & 7.6 & $982.4 \pm 3.0$ & $2.16 \pm 0.01$ \\
 & 1995-07-15 & 8.3 & $1518.8 \pm 7.4$ & $2.46 \pm 0.01$ \\
 & 1997-01-11 & 8.3 & $1150.9 \pm 3.7$ & $2.63 \pm 0.01$ \\
 & 2018-12-04 & 8.7 & $1410.4 \pm 3.0$ & $2.06 \pm 0.01$ \\
J2015$+$3410 & 1996-05-15 & 2.3 & $478.0 \pm 5.2$ & $9.77 \pm 0.13$ \\
 & 2014-06-09 & 2.3 & $518.6 \pm 1.4$ & $8.49 \pm 0.03$ \\
 & 2017-06-15 & 2.3 & $584.1 \pm 2.2$ & $9.63 \pm 0.03$ \\
 & 2018-09-24 & 2.3 & $507.3 \pm 3.2$ & $9.69 \pm 0.06$ \\
 & 1996-05-15 & 8.3 & $643.3 \pm 4.7$ & $0.98 \pm 0.01$ \\
 & 2014-06-09 & 8.7 & $162.1 \pm 0.6$ & $1.06 \pm 0.01$ \\
 & 2017-06-15 & 8.7 & $403.2 \pm 1.0$ & $1.54 \pm 0.01$ \\
 & 2018-09-24 & 8.7 & $304.1 \pm 1.3$ & $1.40 \pm 0.01$ \\
J2015$+$3710 & 2002-01-31 & 2.3 & $1129.8 \pm 5.1$ & $6.48 \pm 0.04$ \\
 & 2014-06-09 & 2.3 & $2006.5 \pm 2.6$ & $7.18 \pm 0.01$ \\
 & 2017-06-10 & 2.3 & $2266.7 \pm 3.2$ & $7.32 \pm 0.01$ \\
 & 2018-09-20 & 2.3 & $1851.5 \pm 4.4$ & $7.19 \pm 0.01$ \\
 & 2016-01-09 & 4.4 & $2758.6 \pm 3.9$ & $2.17 \pm 0.01$ \\
 & 2010-02-13 & 8.4 & $1850.1 \pm 2.8$ & $0.41 \pm 0.01$ \\
 & 2011-08-02 & 8.4 & $2290.8 \pm 1.8$ & $0.56 \pm 0.01$ \\
 & 2002-01-31 & 8.6 & $2368.6 \pm 5.6$ & $0.58 \pm 0.01$ \\
 & 2017-06-10 & 8.7 & $2462.2 \pm 1.3$ & $0.64 \pm 0.01$ \\
 & 2018-09-20 & 8.7 & $2348.6 \pm 2.0$ & $0.52 \pm 0.01$ \\
J2020$+$2942 & 2002-05-14 & 2.3 & $181.0 \pm 1.0$ & $8.04 \pm 0.07$ \\
 & 2014-08-09 & 2.3 & $818.9 \pm 2.4$ & $8.78 \pm 0.03$ \\
 & 2017-07-16 & 2.3 & $727.6 \pm 2.3$ & $8.44 \pm 0.03$ \\
 & 2017-10-21 & 2.3 & $684.3 \pm 2.3$ & $6.90 \pm 0.04$ \\
 & 2018-12-04 & 2.3 & $609.5 \pm 2.3$ & $6.96 \pm 0.04$ \\
 & 2016-06-05 & 4.2 & $432.4 \pm 8.1$ & $8.09 \pm 0.14$ \\ 
 & 2014-08-09 & 8.7 & $98.5 \pm 0.5$ & $0.42 \pm 0.01$ \\
 & 2017-07-16 & 8.7 & $106.4 \pm 0.4$ & $0.28 \pm 0.01$ \\
 & 2017-10-21 & 8.7 & $127.3 \pm 0.7$ & $0.36 \pm 0.01$ \\
 & 2018-12-04 & 8.7 & $108.1 \pm 0.5$ & $0.48 \pm 0.01$ \\
J2023$+$3153 & 2010-07-29 & 1.4 & $1214.7 \pm 1.2$ & $11.77 \pm 0.01$ \\
 & 2000-03-13 & 2.3 & $1506.5 \pm 2.7$ & $5.34 \pm 0.01$ \\
 & 2002-07-24 & 2.3 & $714.4 \pm 3.5$ & $4.55 \pm 0.04$ \\
 & 2006-07-11 & 2.3 & $877.3 \pm 2.9$ & $4.64 \pm 0.02$ \\
 & 2009-01-21 & 2.3 & $1423.4 \pm 2.4$ & $4.35 \pm 0.01$ \\
 & 2013-07-08 & 4.4 & $1219.7 \pm 2.4$ & $1.52 \pm 0.01$ \\
 & 2013-07-08 & 7.6 & $1145.1 \pm 2.7$ & $0.89 \pm 0.01$ \\
 & 1995-10-12 & 8.3 & $1959.1 \pm 2.9$ & $0.64 \pm 0.01$ \\
 & 2000-03-13 & 8.7 & $950.5 \pm 1.4$ & $0.60 \pm 0.01$ \\
 & 2002-07-24 & 8.7 & $491.0 \pm 2.8$ & $0.77 \pm 0.01$ \\
 & 2006-07-11 & 8.7 & $662.9 \pm 1.5$ & $0.54 \pm 0.01$ \\
 & 2008-01-23 & 8.7 & $946.7 \pm 2.7$ & $0.80 \pm 0.01$ \\
 & 2009-01-21 & 8.7 & $1012.4 \pm 2.1$ & $0.96 \pm 0.01$ \\
 
\hline
\end{tabular}
\label{tab:appendix_epochs_part1}
\end{table}

\setcounter{table}{0}
\begin{table}
\centering
\small
\setlength{\tabcolsep}{4pt}
\begin{tabular}{lcccc}
\hline
J2025$+$3343 & 1995-10-12 & 2.3 & $1875.9 \pm 7.9$ & $11.58 \pm 0.04$ \\
 & 1996-05-15 & 2.3 & $224.6 \pm 1.4$ & $7.85 \pm 0.08$ \\
 & 2001-03-12 & 2.3 & $1183.4 \pm 3.6$ & $9.62 \pm 0.03$ \\
 & 2003-07-09 & 2.3 & $665.0 \pm 2.4$ & $8.13 \pm 0.16$ \\
 & 2016-04-19 & 2.3 & $823.2 \pm 6.4$ & $13.13 \pm 0.15$ \\
 & 2018-03-31 & 2.3 & $1621.6 \pm 12.6$ & $16.05 \pm 0.11$ \\
 & 2018-05-30 & 4.4 & $1676.0 \pm 4.3$ & $4.15 \pm 0.01$ \\
 & 2018-05-30 & 7.6 & $1579.5 \pm 2.8$ & $1.13 \pm 0.01$ \\
 & 1995-10-12 & 8.3 & $2628.2 \pm 3.5$ & $0.94 \pm 0.01$ \\
 & 1996-05-15 & 8.3 & $2636.1 \pm 6.2$ & $0.70 \pm 0.01$ \\
 & 2011-08-17 & 8.4 & $2408.9 \pm 2.4$ & $1.23 \pm 0.01$ \\
 & 2016-04-19 & 8.6 & $1082.4 \pm 5.4$ & $1.85 \pm 0.01$ \\
 & 2001-03-12 & 8.7 & $1773.2 \pm 2.3$ & $0.99 \pm 0.01$ \\
 & 2003-07-09 & 8.7 & $1478.0 \pm 2.4$ & $1.60 \pm 0.01$ \\
 & 2018-03-31 & 8.7 & $3045.2 \pm 6.0$ & $0.96 \pm 0.01$ \\
J2050$+$3127 & 1996-05-15 & 2.3 & $826.6 \pm 12.3$ & $10.93 \pm 0.16$ \\
 & 1997-01-11 & 2.3 & $596.3 \pm 4.1$ & $8.98 \pm 0.08$ \\
 & 2005-12-14 & 2.3 & $505.6 \pm 4.2$ & $6.63 \pm 0.23$ \\
 & 2007-01-24 & 2.3 & $581.8 \pm 4.0$ & $8.24 \pm 0.10$ \\
 & 2008-05-14 & 2.3 & $540.7 \pm 3.3$ & $7.70 \pm 0.07$ \\
 & 2017-07-09 & 2.3 & $705.9 \pm 1.9$ & $10.10 \pm 0.03$ \\
 & 2018-07-14 & 4.4 & $483.1 \pm 1.8$ & $3.24 \pm 0.01$ \\
 & 2018-07-14 & 7.6 & $326.1 \pm 1.4$ & $1.18 \pm 0.01$ \\
 & 1996-05-15 & 8.3 & $434.6 \pm 4.8$ & $1.41 \pm 0.02$ \\
 & 1997-01-11 & 8.3 & $403.4 \pm 2.0$ & $0.92 \pm 0.01$ \\
 & 2012-02-20 & 8.4 & $556.2 \pm 1.2$ & $0.88 \pm 0.01$ \\
 & 2005-08-26 & 8.6 & $586.8 \pm 2.4$ & $0.83 \pm 0.01$ \\
 & 2005-12-14 & 8.7 & $547.4 \pm 2.7$ & $0.87 \pm 0.01$ \\
 & 2007-01-24 & 8.7 & $242.9 \pm 1.4$ & $0.94 \pm 0.01$ \\
 & 2008-05-14 & 8.7 & $410.9 \pm 2.5$ & $0.94 \pm 0.01$ \\
 & 2017-07-09 & 8.7 & $425.8 \pm 0.6$ & $0.89 \pm 0.01$ \\
J2223$+$6249 & 2004-05-08 & 2.3 & $156.2 \pm 7.9$ & $10.10 \pm 0.91$ \\
 & 2006-12-18 & 2.3 & $227.2 \pm 4.8$ & $9.76 \pm 0.28$ \\
 & 2017-05-27 & 2.3 & $189.4 \pm 1.3$ & $9.73 \pm 0.07$ \\
 & 2018-07-05 & 2.3 & $176.4 \pm 2.0$ & $9.66 \pm 0.18$ \\
 & 2018-05-26 & 4.4 & $211.9 \pm 0.9$ & $3.15 \pm 0.01$ \\
 & 2018-08-29 & 4.4 & $226.0 \pm 1.0$ & $3.00 \pm 0.02$ \\
 & 2018-05-26 & 7.6 & $182.9 \pm 0.6$ & $1.04 \pm 0.01$ \\
 & 2018-08-29 & 7.6 & $217.0 \pm 0.8$ & $0.98 \pm 0.01$ \\
 & 2012-08-20 & 8.4 & $120.7 \pm 1.7$ & $1.18 \pm 0.02$ \\
 & 2004-05-08 & 8.7 & $143.0 \pm 2.4$ & $0.67 \pm 0.02$ \\
 & 2006-12-18 & 8.7 & $202.3 \pm 3.5$ & $1.01 \pm 0.04$ \\
 & 2017-05-27 & 8.7 & $240.5 \pm 0.4$ & $0.76 \pm 0.01$ \\
 & 2018-07-05 & 8.7 & $220.6 \pm 0.7$ & $0.88 \pm 0.01$ \\
\hline
\end{tabular}
\begin{tablenotes}
    \item Note: $\nu$ is the observing frequency, $S$ and $\theta$ are the flux density and angular size of the AGN, respectively, obtained from a single Gaussian component fit.\\
\end{tablenotes}
\label{tab:appendix_epochs_part2}
\end{table}

\bsp 
\label{lastpage}%
\end{document}